\documentclass[amsmath,amssymb,aps,onecolumn]{revtex4-2}

\usepackage{amsmath, amssymb}
\usepackage{tikz}
\usepackage{caption}
\usepackage{graphics}
\usepackage{xcolor}
\usepackage{bm}
\usepackage{amsthm}
\usepackage{stmaryrd} 
\usepackage{bbold}
\usepackage{caption}
\usepackage{subcaption}
\usepackage{mathrsfs,mathtools}
\usepackage{hyperref}
\usepackage{float}
\hypersetup{
    colorlinks=true,
    linkcolor=blue,
    filecolor=blue,      
    urlcolor=blue,
    citecolor=blue
  }
\usepackage{xr}

\newcommand{\1}{\mathbb{1}}
\newcommand{\Tr}{\text{Tr}}

\renewcommand{\d}{\textrm{d}}

\begin{document}

\title[Jamming pair of general run-and-tumble particles]{Jamming pair of general run-and-tumble particles:\\ Exact results, symmetries and steady-state universality classes}

\author{Leo Hahn$^1$, Arnaud Guillin$^1$, Manon Michel$^1$}
\email{manon.michel@cnrs.fr}

\address{$^1$ Laboratoire de Math\'ematiques Blaise Pascal UMR 6620, CNRS, Universit\'e Clermont-Auvergne, Aubi\`ere, France}

\begin{abstract}
  While run-and-tumble particles are a foundational model for
  self-propelled particles as bacteria or Janus particles, the
  analytical derivation of their steady state from the microscopic
  details is still an open problem. By directly modeling the system at
  the continuous-space and -time level thanks to piecewise
  deterministic Markov processes (PDMP), we derive the conservation
  conditions which sets the invariant distribution and, more
  importantly, explicitly construct the two universality classes for
  the steady state, the \emph{detailed-jamming} and the
  \emph{global-jamming} classes. They respectively identify with the
  preservation or not in a detailed manner of a symmetry at the level
  of the dynamical internal states between probability flows entering
  and exiting jamming configurations.  We call such symmetry active
  global balance, as it is the true nonequilibrium counterpart of the
  equilibrium global balance. Thanks to a spectral analysis of the tumble kernel,
  we give explicit expressions for the invariant measure in the
  general case. We show that the non-equilibrium features exhibited by the steady
  state include positive mass for the jammed configurations and, for
  the global-jamming class, exponential decay and growth terms,
  potentially modulated by polynomial terms. Interestingly, we find that the
  invariant measure follows, away from jamming configurations, a
  catenary-like constraint, which results from the interplay between
  probability conservation and the dynamical skewness introduced by
  the jamming interactions, seen now as a boundary constraint.  This
  work shows the powerful analytical approach PDMP provide for the
  study of the stationary behaviors of RTP systems and motivates their
  future applications to larger systems, with the goal to derive
  microscopic conditions for motility-induced phase transitions.
 
\end{abstract}

\maketitle

\section{Introduction}

Active matter systems such as bacterial colonies \cite{schnitzer93,
  wu00, cates12, elgeti15} or robot swarms \cite{mayya19, wang21} are
characterized by the breaking at the microscopic scale of energy
conservation and time reversibility~\cite{ramaswamy10,
  marchetti2013,obyrne22}. Compared to their passive counterparts,
these systems exhibit rich behaviors including motility-induced phase
separation \cite{cates2015, caprini20}, emergence of patterns
\cite{surrey2001} or collective motion \cite{vicsek12,
  bricard2013,keta2021}. Their study remains challenging because many tools
from equilibrium statistical mechanics do not apply. In particular
there is no guarantee the steady state follows some Boltzmann
distribution, even up to the definition of an effective
potential. This has fueled interesting developments in both physics
and mathematics, but which typically rely on a coarse-grained approach
\cite{toner98, marchetti2013, tailleur08, farage15, steffenoni17,
  laighleis18}. While they successfully recover some out-of-equilibrium
macroscopic behaviors, they leave open the question of the microscopic
origin of a given phenomenon, a question strongly linked to the
fundamental one of universality.

It motivates the derivation of an exact theory of active particles
from the microscopic scale, without any coarse-graining apart the
stochastic description of individual particle motion. Recent works
\cite{kourbane18, erignoux21} deal with the derivation of large-scale
limit of active lattice-gas models \cite{kipnis98}, typically aiming
at deriving the steady state. While being exact, such hydrodynamic
limit requires strong conditions (symmetric jumps, particular
exclusion rule), in order to ensure a rigorous limit
derivation. Furthermore, any results at a large but finite size
requires a subtle computation of fluctuating terms, as done in
Macroscopic Fluctuation Theory \cite{derrida07,bertini02}. Whereas
such terms impacts the existence of metastable solutions, their
derivation remains challenging \cite{agranov21} and is more often
analytically or numerically approximated
\cite{chatterjee11,dolai20,chakra1,chakra2}.

Another approach focuses on the derivation of exact results for
microscopic models \cite{slowman2016, slowman2017}, with a strong
motivation to decipher the impact of the reversibility breaking. They
consider discrete models of two Run-and-Tumble Particles (RTP) with a
hardcore jamming interaction on a 1D-ring. The RTP is a paradigmatic
active matter model mimicking the moving pattern of bacteria such as
E. Coli \cite{schnitzer93}. RTPs perform a series of straight runs
separated by stochastic refreshments of the run direction called
tumbles. Even a single RTP displays interesting out-of-equilibrium
features \cite{malakar18, dhar19, ledoussal19, basu20, mori21}. In
\cite{slowman2016,slowman2017}, the steady state distribution on a
lattice is exactly determined from a Master equation by a generating
function approach, respectively for the case of instantaneous and
finite tumbles. The continuous limit of such distribution is also
discussed. They identify the Dirac jamming contributions to the
stationary distribution and a new lengthscale in the finite tumble
case, linked to an exponential decay or growth. While these results
are a first, questions remain open regarding the rigorous continuum
limit of such lattice models, the portability of such involved methods
to more general systems and dimensions and the actual capacity of
lattice models to capture the particular role that domain boundaries
seem to play in such piecewise-ballistic systems. An attempt to
directly study these systems in continuous space has already been
made, but with some additional thermal noise so that the analysis
of boundary conditions is more tractable \cite{das20}. They recover
the findings of \cite{slowman2016} for the instantaneous tumble case,
plus some additional exponential decays due to the thermal
noise. However, the rigorous derivation of the vanishing noise limit
and its generalization to other tumble scenarios remain open.  More
importantly, the definition of some general universality classes
ruling the steady state remains yet open.

In this work, we address the universality question for
  a pair of RTPs with jamming interactions and a general tumble
  mechanism, i.e. there is no assumption on the internal velocity
  space size or structure. We show that such general two-RTP system
  with jamming is entirely described by two explicit universality
  classes we name the detailed-jamming and global-jamming
  classes. Such classes identify with the detailed preservation or not
  of a global dynamical symmetry at jamming. This dynamical symmetry,
  we name active global balance, directly replaces the time
  reversibility one and is the true nonequilibrium counterpart of the
  equilibrium global balance. We then derive explicit steady-state
  expression of the different classes. The key idea is to build a
  formalism of the RTP dynamics incorporating all the present
  symmetries (particle indistinguishability, space homogeneity) and
such directly at the continuous-space and -time level thanks to
piecewise deterministic Markov processes (PDMP) \cite{davis84,
  davis93}. We then gain direct access to the impact of any dynamical
skewness introduced by jamming. We show that it takes the form of a
boundary constraint on the bulk dynamical relaxation, that is ruled by
the tumbles. We more precisely demonstrate that the nonequilibrium
features always includes positive mass for the jammed configurations
but exponential decay and growth terms only when the detailed
dynamical symmetry is broken. These effective attractive forces and
respective lengthscales can be understood as stemming from a
catenary-like interplay between probability conservation and any
dynamical skewness forced by jamming. Therefore, this formalism based
on a PDMP characterization highlights the role played by boundaries
and leads to the identification of key fundamental symmetries, which
could lead to important advances in understanding MIPS.

Thus, we provide in Section~\ref{sec:describe} a description of the
general two-RTP process as PDMP, including their generator (an
infinitesimal description of the evolution). This is the basis to
obtain in Section~\ref{sec:system} the conservation constraints on the
steady state distribution and in Section~\ref{sec:classes} the
explicit definition of the two universality detailed-jamming and
global-jamming classes and their respective steady-state
expressions. They are consistent with the continuous-space limit of
the particular discrete cases in \cite{slowman2016,slowman2017} or the
particular continuous system in \cite{keta2021}. We furthermore derive
the explicit expressions of the steady-state for general two-state RTP
pairs, forming up a detailed-jamming subclass, in
Section~\ref{sec:app-2rtp}. In Section~\ref{sec:app-3rtp-gen}, we do
the same for general isotropic and anisotropic three-state RTP, which
are important global-jamming subclasses. This led us to clarify the
role played by the relaxation lengthscale into the separation of some
diffusive and ballistic regimes. Finally we conclude in
Section~\ref{sec:discuss} with a discussion on the light now shed on
the interplay between reversibility breaking and jamming and the
perspectives it brings.

 \section{Two-RTP system as PDMP}
  \label{sec:describe}
\begin{figure}
\centering
\captionsetup{font=small}
\includegraphics[width=1.0\textwidth]{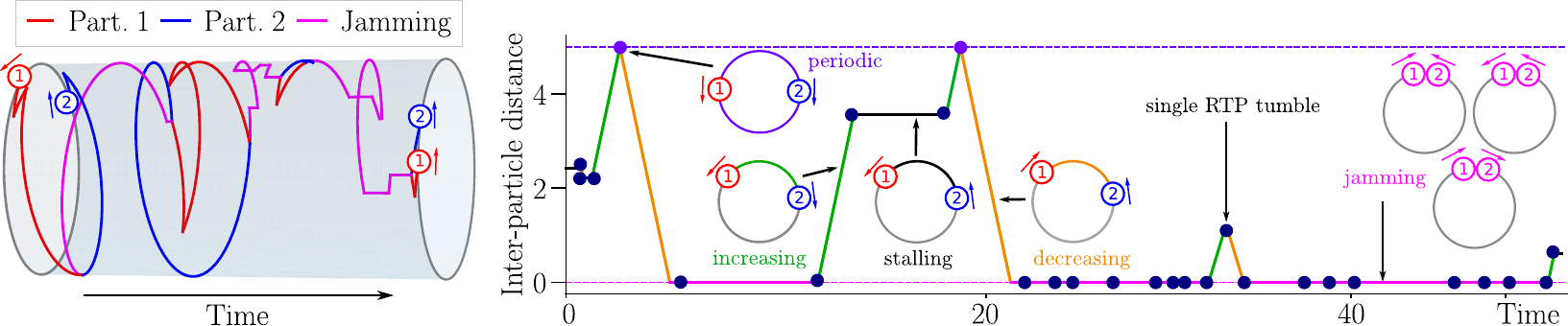}
\caption{Trajectory of two RTPs ({\bf left}) undergoing instantaneous
  tumble ($\omega = 0.2$, $L=10$), and of their corresponding periodic
  interdistance ({\bf right}), which evolution undergoes several regimes in the
  bulk (increasing, stalling and decreasing (\textbf{a})) and boundary conditions
  (periodic at $L/2$ (\textbf{b}) and jamming at $0$ (\textbf{c})).}
\label{fig:trajectory}
\end{figure}

  We consider a pair of interacting RTPs on a one-dimensional torus of
  length $L$. A single-particle possible internal state is described
  by the variable $v$, which typically codes for the velocity of the
  particle and commonly takes values in $\{-1,+1\}$ (two-state RTP or
  instantaneous tumble) or in $\{-1,0,+1\}$ (three-state RTP or finite
  tumble) as in \cite{slowman2016,slowman2017}. Aiming at universality
  class definition, we adopt a general setting and, therefore, we do
  not restrict the value of $v$, which can be continuous as in
  Section~\ref{sec:app-2rtp}, or anistropic as in Section
  \ref{sec:app-anisotropic}. Both particles change their internal
  state independently according to a Poisson jump process set by the
  transition rate $\omega(v)>0$ and Markov kernel $q$, which are general but
  homogeneous in space and yielding ergodicity in the internal-state
  space. The ergodicity assumption does not restrict generality, as it
  is always possible to consider the ergodic subsystem. The particles interact with each
  other through hardcore interactions leading to jammed states where
  the two RTPs are colliding against each other until a tumble allows
  the particles to escape, as illustrated in
  Fig.~\ref{fig:trajectory}.

  We now show how to formalize the system and all its
  symmetries in a continuous-time and continuous-space setting by
  using PDMP.  First, we note $x_{t,1},x_{t,2}$ the
  positions of the particles, and $v_{t,1}$ and $v_{t,2}$ the particle
  internal states. Thanks to the homogeneity, periodicity and particle
  indistinguishability, the system can entirely be
  described at time $t$ by the periodic interdistance
  $0\leq r_t \leq L/2$ between particles, defined as
  $\min(|x_{t,1}-x_{t,2}|, L-|x_{t,1}-x_{t,2}|)$. As shown in
  Fig.~\ref{fig:trajectory}, the interdistance $r_t$ then
  undergoes the following evolution, depending if it is located in the
  \emph{bulk} (a), at the \emph{periodic} (b) or \emph{jamming} (c)
  boundaries:

  {\bf(a - Run)} While in the \emph{bulk} ($0<r_t<L/2$), the interdistance $r_t$
  is updated through a deterministic evolution in three possible
  regimes: increasing ($(v_{t,1}-v_{t,2})(x_{t,1}-x_{t,2}) \gtrless 0$
  if $|x_{t,1}-x_{t,2}|\lessgtr L/2$), decreasing
  ($ (v_{t,1}-v_{t,2})(x_{t,1}-x_{t,2}) \lessgtr 0$ if
  $|x_{t,1}-x_{t,2}|\lessgtr L/2$) and stalling ($v_{t,1}=v_{t,2}$).

  {\bf(a - Tumble)} The transitions inside and between regimes stem
  from a particle tumble and are ruled by the superposed
  Poisson process of rate $\omega(v_{1}) + \omega(v_{2})$. No direct
  transition between two states of the stalling regime can occur, as
  it would require a simultaneous tumble from both particles, which is
  of null probability.

  {\bf(b)} When reaching the \emph{periodic} boundary ($r_t = L/2$) in
  an increasing regime, the interdistance automatically switches, by
  periodicity, from an increasing to a decreasing regime. 

  {\bf(c)} Finally, and most interestingly, when the \emph{jamming}
  boundary ($r_t = 0$) is reached in a decreasing regime, the
  interdistance is stalling in $0$ until a tumble occurs allowing the
  particles to separate from each other and making $r_t$ re-enter the
  \emph{bulk} in an increasing regime.

We now describe the stochastic dynamics of $r_t$ by a PDMP
\cite{davis84,davis93}. A PDMP $(X_t,V_t)$, with $X$ some positional
and $V$ some dynamical variables, is entirely encoded by its
generator. Roughly speaking, the extended generator $\mathcal A$,
applied to a suitable class of function (including boundary
conditions), is such that
$$ f(X_t,V_t)-f(X_0,V_0)-\int_0^t\mathcal{A}f(X_u,V_u)\d u$$
is a local martingale, and may be seen unformally as
$$\mathcal{A}f(x,v)=\lim_{t\to0^+}\frac {E_{(x, v)}[f(X_t, V_t)]-f(x,v)}t.$$
Unformally, the semigroup
$P_tf(x,v):=E_{(x,v)}[f(X_t,V_t)]$, may be rewritten as
$P_t=e^{t\mathcal A}$.  We refer to \cite[Sec. 26]{davis93}
for a systematic treatment of the extended generator of a PDMP. The
explicit expression of the generator is a powerful tool to study the
longtime behavior of a Markov process, for example to compute the
invariant measure(s) or assess convergence to equilibrium via Lyapunov
conditions \cite{MT93}.

\begin{figure}
\centering
\captionsetup{font=small}
\includegraphics{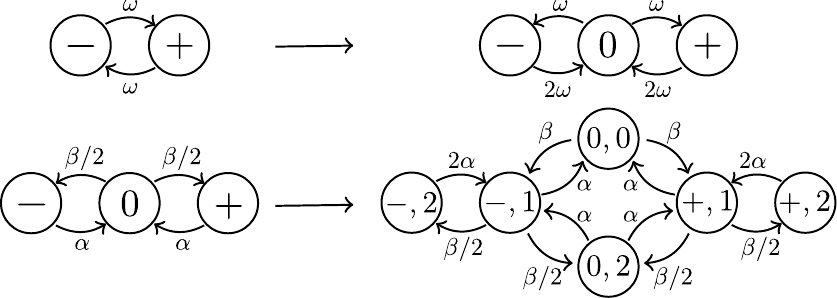}
\caption{Transitions between the internal states at the single
  particle ($\omega(\cdot)q(\cdot,\cdot)$, {\bf left}) and
  interdistance level ({$\tilde{\omega}(\cdot)Q(\cdot,\cdot)$, \bf
    right}) in the case of instantaneous ({\bf top}, and finite tumble
  ({\bf left}), in a reduced representation for $s$ thanks to the
  single-particle tumble isotropy (instantaneous: $s$ can be omitted,
  finite: $s$ codes for the number of moving particles).}
\label{fig:transition_rates}
\end{figure}

Starting from the previous description of the process, and
following \cite{davis93}, we now characterize the PDMP
ruling the evolution of $(r_t,\nu_t,s_t)$:

{\bf(a)} by its infinitesimal
generator in the bulk, $0<r_t<L/2$, 
\begin{equation*}
    \mathcal{A}_{\mathcal{B}}\!=\! \underbrace{\nu\phi(s)\partial_r}_{\text{Run phase}}  + \underbrace{\tilde{\omega}(s) (Q((\nu,s),\cdot)-\text{Id})}_{\text{Tumble phase}},
  \label{eq:generator}
\end{equation*}
where we introduce the two auxiliary dynamical variables
$(\nu_t,s_t)\in\mathcal{V}$, so that the complete bulk set is
$\mathcal{B}=]0,L/2[\times\mathcal{V}$: the variable
$\nu_t \in\{-1,0,1\}$ sets the nature of the regime as defined in (a)
and $\mathcal{V}$ can be decomposed as
$\mathcal{V}_{-1}\cup \mathcal{V}_0\cup\mathcal{V}_1$. The multiset variable
$s_t=\{\{v_{t,1},v_{t,2}\}\}$, encoding the particle indistinguishability,
sets the amplitude of the evolution by $\phi(s_t)=|v_{t,1}-v_{t,2}|$
and the tumble by its rate
$\tilde{\omega}(s_t)=\omega(v_{t,1})+\omega(v_{t,2}) $ and its
transition Markov kernel $Q((\nu,s), \cdot)$,
\begin{equation}
  \begin{aligned}
    Q((\nu,\{v,\tilde{v}\}), (\nu',\{v',\tilde{v}\}))= & q(v,v')\Big(\tfrac{1}{2}\1_{\{0\}}(\nu)\1_{\{-1,1\}}(\nu')
  &+\tfrac{\omega(v)}{\tilde{\omega}(\{v,\tilde{v}\})}\1_{\{-1,1\}}(\nu)\1_{\{\text{sign}(\nu\nu')\}}\big(\text{sign}\big(\tfrac{v'-\tilde{v}}{v-\tilde{v}}\big)\big)\Big) 
,
\end{aligned}
\end{equation}
all other transitions being of probability 0, as they would involve
tumble of both particles. The transition rates at the single-particle
($\omega(\cdot)q(\cdot,\cdot)$) and interdistance
($\tilde{\omega}(\cdot)Q(\cdot,\cdot)$) level are shown in
Fig.~\ref{fig:transition_rates} for the instantaneous and finite
tumble cases.

Most importantly, the particle indistinguishability leads to an
isotropic symmetry for $Q$,
\begin{equation}
  Q((\nu,s), (\nu', s')) = Q((-\nu,s), (-\nu', s')),
\label{eq:sym_Q}
\end{equation}
as illustrated on Fig.~\ref{fig:transition_rates} for the
instantaneous and finite tumbles. The
symmetry (\ref{eq:sym_Q}) will have an important impact on the
possible invariant distributions and justifies the introduction of the
representation $(\nu,s)$.

{\bf (b)} by its behavior at the periodic boundary: at
$(r,\nu,s)\in\{L/2\}\times\mathcal{V}_{1}=\Gamma^*_P$,
the periodic boundary kernel
$$ Q_P((r,\nu,s), (r',\nu', s')) = \1_{\{(r,-\nu,s)\}}(r',\nu',s')$$
  codes for a switch from an increasing ($\nu=1$) to a
  decreasing ($\nu'=-1$) regime.
  
 {\bf(c)} and finally by its behavior at the jamming boundary: for
  $(r,\nu,s)\in \{0\}\times\mathcal{V}_{-1} = \Gamma^*_J$, the
  jamming boundary kernel, $$Q_J((r,\nu,s),( r^*,\nu^*,  s^*))
    = \1_{\{(r,\nu,s)\}}(r^*,\nu^*,s^*)$$
  makes the system jump into a jammed configuration
  $(r^*,\nu^*,s^*)\in
  \mathcal{J}=\{0\}\times\mathcal{V}_{-1}\cup\mathcal{V}_{0} $, which is ruled
  by the following infinitesimal generator,
  \begin{equation}
      \mathcal{A}_{\mathcal{J}} \!=\!  \tilde{\omega}(s^*) (K((\nu^*,s^*), \cdot )- \text{Id}).
  \label{eq:jamming_gen}
\end{equation}
It translates the persistence of the jamming state until the tumbles
lead to an unjamming state ($\nu=+1$). There, it generates an
increasing regime in the bulk, once rigorously combined with an
unjamming boundary kernel analogous to $Q_{\mathcal{J}}$. As most of
the results derived in the following Sections are general, we
introduce here a jamming tumble Markov kernel $K$ that may be different from the bulk
tumble kernel $Q$. In most cases of interest, $K$ identifies with $Q$
and we further analyze this situation.

\section{Conservation constraints in the steady state and jamming
  boundary role}
\label{sec:system}

The steady-state distribution of the periodic interdistance $r$ of two
passive hardcore particles is the uniform density over $[0,L/2]$, with
in particular a null probability to find $r$ in the exact state
$0$. The addition of nonequilibrium features (activity, jamming) may
introduce deviations from this equilibrium behavior, which we now
study. In this Section, we derive the necessary and sufficient
conservation constraints on the probability flows in the steady state
and define the \emph{active} global and detailed balances,
nonequilibrium counterparts of global and detailed balances.

The measure $\pi$ left invariant by the considered PDMP should satisfy
the following condition,
  \begin{equation}\label{eq:app_invariance}
  \int_{\mathcal{B}} \mathcal{A}_Bf \d\pi +  \int_{\mathcal{J}} \mathcal{A}_Jf\d\pi = 0,
 \end{equation}
 with $f$ a suitably smooth test function (here we do not detail the
description of the extended domain of the generator and a suitable
core, as required by \cite[Th. 5.5]{davis93}), satisfying in particular the periodic boundary condition. As
 the considered process is trivially irreducible thanks to the
 ergodicity of the tumbles, the distribution $\pi$ is unique. Stemming from the periodic boundary kernel $Q_P$, $f$ shall verify
the condition 
\begin{equation*}
f(L/2,\nu,s) = f(L/2,-\nu,s).
\end{equation*}
Similar conditions at the jamming boundary are required, i.e.
\begin{equation*}
f(0^+,\nu,s) = f(0,\nu,s),   
\end{equation*}
where we omit out of simplicity the superscript $^*$ characterizing
the jamming state. Now, expliciting the bulk term in
\eqref{eq:app_invariance},
\begin{multline*}
\int_{\mathcal{V}_{-1}\cup\mathcal{V}_0} \mathcal{A}_Jf(0,\nu,s)\pi(0,d\nu,d s)
+\int_{]0,L/2[\times\mathcal{V}}\d\pi(r,\nu, s) \nu\phi(s)\partial_rf(r,\nu,s)\\
+\int_{]0,L/2[\times\mathcal{V}}\d\pi(r,\nu, s) \tilde{\omega}(s)
\int_{\mathcal{V}} Q((\nu,s),(\d\nu'\d s'))\Big(f(r,\nu',s') - f(r,\nu,s)\Big)        
= 0,
\end{multline*}
we obtain by integration by parts in the second term,
\begin{equation*}
\begin{split}
0&=\int_{\mathcal{V}_{-1}\cup\mathcal{V}_0} \mathcal{A}_Jf(0,\nu,s)\pi(0,\d\nu,\d s)
- \int_{\mathcal{V}_{-1}\cup\mathcal{V}_{1}}  \nu\phi(s)\pi(0^+,\d\nu,\d s)f(0,\nu,s) + \int_{\mathcal{V}_{-1}\cup\mathcal{V}_{1}}  \nu\phi(s)\pi(L/2^-,\d\nu,\d s)f(L/2,\nu,s) \\
&+\int_{]0,L/2[\times\mathcal{V}}\d r\d\nu\d sf(r,\nu,s)
\Big[\int_{\mathcal{V}}\pi(r,\d\nu',\d s')\tilde{\omega}(s')Q((\nu',s'),(\nu, s))  - \pi(r,\nu,s)\tilde{\omega}(s)  - \nu\phi(s)\partial_r\pi(r,\nu,s)  \Big].
\end{split}
\end{equation*}As $f(L/2,\nu,s) = f(L/2,-\nu,s)$ by the condition on the
periodic boundary and as $\pi(L/2,-\nu, s)=\pi(L/2,\nu, s)$ by
definition of the periodic interdistance, we have the
cancellation of the term in $L/2$ by periodicity,
\begin{equation}
\int_{\mathcal{V}_{1}}  \nu\phi(s)\pi(L/2,\d\nu,\d s)f(L/2,\nu,s) +  
\int_{\mathcal{V}_{-1}}  -(-\nu)\phi(s)\pi(L/2,-\d\nu,\d s)f(L/2,-\nu,s) = 0.
\label{eq:periodic_boundary_cancel}
\end{equation}
Thus we have, expliciting the jamming generator,
\begin{equation*}
\begin{split}
&0=\int_{\mathcal{V}} \d\nu \d sf(0,\nu,s) \Big[\int_{\mathcal{V}_{-1}\cup\mathcal{V}_{0}}\pi(0,\d\nu',\d s')\tilde{\omega}(s') K((\nu',s'),(\nu,s))- \1_{\{-1,0\}}(\nu)\pi(0,\nu,s)\tilde{\omega}(s)
-  \nu\phi(s)\pi(0^+,\nu,s)\Big] \\
&\quad+\int_{]0,L/2[\times\mathcal{V}}\d r\d\nu \d s f(r,\nu,s) \Big[\int_{\mathcal{V}}\pi(r,\d\nu',\d s')\tilde{\omega}(s') K((\nu',s'),(\nu,s))- \pi(r,\nu,s)\tilde{\omega}(s) - \nu\phi(s)\partial_r\pi(r,\nu,s))\Big].
\end{split}
\end{equation*}
As $f$ can be any test function and recalling the periodic condition, we obtain
for any $\nu,\ s$ the following condition system equivalent to \eqref{eq:app_invariance},
    \begin{equation}
\def\arraystretch{1.4}
      \left\{\begin{array}{l}
        (\textrm{C}_{\mathcal{B}}) \ \int_{\mathcal{V}}\pi(r,\d\nu'\d s')\tilde{\omega}(s') Q((\nu',s'),(\nu, s)) \!  =\!   \tilde{\omega}(s)\pi(r, \nu, s) + \nu\phi(s)\partial_r\pi(r,\nu,s)
       \\         
        (\textrm{C}_P) \  \pi(L/2,1,s)  =\pi(L/2,-1,s)\\
         (\textrm{C}_{\mathcal{J}}) \  \int_{\mathcal{J}}\d\pi(0,\nu',s')\tilde{\omega}(s')K((\nu',s'),(\nu,s))
          =  \1_{\{-1,0\}}(\nu)\tilde{\omega}(s)\pi(0, \nu, s) + \nu\phi(s)\pi(0^+,\nu,s).
      \end{array}\right.
    \label{eq:conditions}
  \end{equation}

    This condition system completely constrains the invariant
    measure $\pi$. It can be understood as a set of conservation
    equations of probability flows in the bulk $\mathcal{B}$, and on
    the jamming and periodic boundaries. It leads to the
    following conclusions:\medskip
    
    (i) $(\textrm{C}_P)$ is a direct translation of the periodicity of the
    system and is not impacting the physical meaning of the form of
    $\pi$, in particular potential deviation from the passive
    equilibrium form.\medskip
    
    (ii) In the bulk, $(\textrm{C}_{\mathcal{B}})$ is setting a
    balance between the Run and Tumble contributions. We define the
    \emph{tumble} stationary measure $\mu_{\mathcal{B}}(\cdot)$ as the unique
    probability measure so that
    $\tilde{\omega}(\cdot)\mu_{\mathcal{B}}(\cdot)$ is left invariant
    by the tumble kernel $Q$, i.e.    
\begin{equation}
  \int_{\mathcal{V}}\mu_{\mathcal{B}}(\d\nu'\d s')\tilde{\omega}(s') Q((\nu',s'),\d\nu\d s)
  =   \mu_{\mathcal{B}}(\d\nu, \d s)\tilde{\omega}(s).
  \label{eq:mub_cond}
\end{equation}
Any product measure of the form
    $\tfrac{2}{L}\1_{]0,L/2[}(r)\mu_{\mathcal{B}}(\cdot)$,  is
    satisfying $(\textrm{C}_\mathcal B)$ with a null Run term
    $\phi(\cdot)\partial_r\pi(r,\pm 1,\cdot)\!=\!0$.
    Such a product form corresponds to the passive counterpart.

    Furthermore, the
    isotropy of $Q$ (\ref{eq:sym_Q}) leads to
    $\mu_{\mathcal{B}}(\nu,s)=\mu_{\mathcal{B}}(-\nu,s)$, satisfying
    $(\textrm{C}_P)$. Reciprocally, any distribution such that
    $\pi(r,\nu,s)=\pi(r,-\nu,s)$ is obeying $(\textrm{C}_\mathcal B)$ only if
    $\partial_r\pi=0$ for all $r>0$, i.e. only if
    $\tilde{\omega}(\cdot)\pi(\cdot)$ is invariant by $Q$ and then identifies
    with $\tilde{\omega}(\cdot)\mu_{\mathcal{B}}(\cdot)$.

    \medskip
    
    (iii) The jamming condition $(\textrm{C}_J)$
    introduces for finite $\omega$ Dirac delta
      components into the invariant measure $\pi$, indicating finite
      probability masses at $r=0$ (jamming configurations). Unlike in
      the passive case, the system under jamming spends a finite
      amount of time in these configurations before unjamming,
      resulting in a finite probability of occupying them despite the
      continuous nature of the state space.\medskip

      (iv) It appears clearly that the jamming
      interaction can impact the bulk behavior by the term
      $\phi(s)\pi(0^+,\pm 1,s)$ in $(\textrm{C}_{\mathcal{J}})$. It
      can be understood as a boundary source term for the bulk
      equation set by $(\textrm{C}_{\mathcal{B}})$, which can force
      the system away from the uniform
      distribution in $r$ in the bulk. This term comes from the
      integration by part, which stems from the Run term in
      $\mathcal{A}_B$. It is indeed the
      activity of the particles that pushes the exploration up to the
      boundary of the state space.\medskip

      (v) We can derive the nonequilibrium counterpart of global
      balance, i.e. an \emph{active} \emph{global} balance
      condition. Indeed, $(\textrm{C}_{\mathcal{J}})$ yields the
       balance of the probability flows in and out of jamming,
      $ \int \d s \phi(s)\pi(0^+,-1,s) = \int \d s
      \phi(s)\pi(0^+,1,s)$ and $(\textrm{C}_\mathcal B)$ imposes the 
      balance at any $r>0$
      $ \int \d s \phi(s)\partial_r\pi(r,-1,s) = \int \d s
      \phi(s)\partial_r\pi(r,1,s)$. Both combined, we obtain the
      following \emph{active} \emph{global} balance of the probability
      flows at any $r>0$,
    \begin{equation}
      \int \d s \phi(s)\pi(r,-1,s) =  \int \d s \phi(s)\pi(r,1,s).
      \label{eq:all_flow}
    \end{equation}
    From (ii), any distribution $\pi$ obeying the stricter
    \emph{detailed} counterpart of (\ref{eq:all_flow}), i.e., for
    any $r, s$,
    \begin{equation}
      \pi(r,-1,s) =  \pi(r,1,s),
      \label{eq:det_flow}
    \end{equation}
    must recover with the tumble stationary distribution in the bulk,
    i.e.
    $$ \pi(r,-1,s) =  \pi(r,1,s) = \tfrac{2}{L}\1_{]0,L/2[}(r)\mu_{\mathcal{B}}(\pm 1,s).$$ Actually, any distribution $\pi$
    obeying (\ref{eq:det_flow}) in at least a single point $r_0<L/2$
    satisfies (\ref{eq:det_flow}) in all $r$.  Indeed,
      the existence of such a point $r_0$ imposes the distribution to
      be uniform in $r$ for $r_0\leq r \leq L/2$, which, given
      $C_{\mathcal{B}}$, imposes uniformity for all $r$.  Therefore,
    while time reversibility and its detailed balance is broken, the
    preservation of such detailed dynamical symmetry restores in the
    bulk the uniform distribution, which is the passive stationary
    state.\medskip

    Thus, analyzing the conservation conditions in
    (\ref{eq:conditions}) unravels the potential impact of the
    nonequilibrium features on the invariant measure: First, the
    presence of Dirac terms in $r=0$ for finite $\omega$; Second, the
    deviation from a product measure uniform in $r$ in the bulk for
    unjamming scenarios which sets boundary conditions
    $\pi(0^+,\cdot)$ different from the tumble stationary distribution
    $\mu_{\mathcal{B}}(\cdot)$, i.e. breaking the detailed symmetry \eqref{eq:det_flow}.

      \section{Universality classes and expression of the steady state}
      \label{sec:classes}
      
      We identify two universality classes, based on the
      deviation or not of $\pi(0^+,\pm 1,s)$ from
      $\tfrac{2}{L}\1_{]0,L/2[}(r)\mu_{\mathcal{B}}(\pm 1, s)$. Such deviation can always be explicitly
      tested in $(C_{\mathcal{J}})$. It is equivalent to the
      satisfaction or breaking of the detailed symmetry upon entering and
      exiting jamming for any $s$,
      \begin{equation}
        \pi(0^+,1,s) = \pi(0^+,-1, s).
        \label{eq:DB}
      \end{equation}
      
      \subsection{Detailed-jamming class}
      The detailed symmetry (\ref{eq:DB}) is satisfied. As a
      consequence, the active global balance (\ref{eq:all_flow})
      condition is obeyed through the active detailed one
      (\ref{eq:det_flow}). Furthermore the nonequilibrium nature is
      only apparent through the presence of Dirac terms at the jamming
      point in the invariant distribution
      $\pi_{\textrm{det}}(r,\nu,s)$, which is of the form,
      \begin{equation}
        \pi_{\textrm{det}}(r,\nu,s) = w_{\mathcal{J}}\1_{\{0\}}(r)\mu_{\mathcal{J}}(\nu,s) + w_{\mathcal{B}}\tfrac{2}{L}\1_{]0,L/2[}(r)\mu_{\mathcal{B}}(\nu,s),
        \label{eq:pi_detjam}
      \end{equation}
      with $\mu_{\mathcal{J}}$ the invariant jamming distribution and
      $w_{\mathcal{J}}, w_{\mathcal{B}}$ the average time ratio
      respectively spent in jamming and the bulk. Very close to an
      equilibrium counterpart, it follows in the bulk a product form,
      which, once marginalized over the dynamical variables,
      identifies with the passive uniform steady-state.  In
      particular, this is the case of the instantaneous tumble
      \cite{slowman2016} but also of any systems with only two
      internal states $\pm v$.

      We now determine in more details the steady-state
      distribution. Plugging the detailed-jamming symmetry
      (\ref{eq:DB}) into $(\textrm{C}_J)$ (\ref{eq:conditions}), we
      obtain,
      \begin{align}
        \int_{\mathcal{V}_{-1}\cup\mathcal{V}_{0}}w_{\mathcal{J}}\d\mu_{\mathcal{J}}(\nu',s')\tilde{\omega}(s')K((\nu',s'),(1, s))&=\phi(s)\pi(0^+,1,s)= -(-1)\phi(s)\pi(0^+,-1,s),
        \label{eq:muj-+-cond}
      \end{align}
      so that,
      \begin{equation}      
        \int_{\mathcal{V}_{-1}\cup\mathcal{V}_{0}}\d\mu_{\mathcal{J}}(\nu',s')\tilde{\omega}(s')[K((\nu',s'),(-1, s)) + K((\nu',s'),(1, s))- \delta(\nu'+1)\delta(s'-s)] = 0.
       \label{eq:muj-1-cond}
        \end{equation}
        From $(\textrm{C}_J)$, the condition for $\nu=0$ writes, as
        $K((0,s'),(0, s))=0$ for all $s,s'$ (no simultaneous double
        tumbles),
       \begin{equation}
\int_{\mathcal{V}_{-1}}\d\mu_{\mathcal{J}}(\nu',s')\tilde{\omega}(s')K((\nu',s'),(0, s))
\!=\! \tilde{\omega}(s)\mu_{\mathcal{J}}(0,s).
\label{eq:muj-0-cond}
\end{equation}
The jamming invariant measure $\mu_{\mathcal{J}}$ is then completely
characterized by both previous conditions and only depends on the
unjamming scenario set by $K$. 
  Finally, the ratio $w_\mathcal{J}/w_\mathcal{B}$ of time spent
  respectively while jamming and in the bulk is set
  by injecting the expression of $\mu_{\mathcal{J}}$ and
  $\pi(0^+,1,s)$ in \eqref{eq:muj-+-cond},
  \begin{equation}
    \int_{\mathcal{V}_{-1}\cup\mathcal{V}_0}\d\mu_{\mathcal{J}}(\nu',s')\tilde{\omega}(s')K((\nu',s'),(1, s)) =  \frac{2\phi(s)}{L}\frac{w_{\mathcal{B}}}{w_{\mathcal{J}}}\mu_{\mathcal{B}}(1,s),
    \label{eq:wj-wb-cond}
  \end{equation}
  
  {\noindent \textbf{ Case $K=Q$.}} We now show that for $K=Q$ in the
  detailed-jamming class, the full jamming distribution
  $\mu_{\mathcal{J}}$ actually identifies with the tumble one. Indeed,
  using \eqref{eq:muj-0-cond} in \eqref{eq:muj-1-cond} and the
  symmetry \eqref{eq:sym_Q} of $Q$, we obtain,
\begin{multline}
\int_{\mathcal{V}_{-1}}\d\mu_{\mathcal{J}}(\nu',s')\tilde{\omega}(s')\Big[Q((\nu',s'),(-1, s)) + Q((\nu',s'),(1, s))   - \delta(s'-s)
+2\int_{\mathcal{V}_0} \d \nu''\d s''Q((\nu',s'),(0, s'')) Q((0,s''),(-1, s)) \Big] \\=0.
\label{eq:muj-det-cond}
  \end{multline}
  Interestingly, the invariance of the tumble distribution
  \eqref{eq:mub_cond} yields the same condition
  \eqref{eq:muj-det-cond} thanks to the detailed symmetry of
  $\mu_{\mathcal{B}}(1,s)=\mu_{\mathcal{B}}(-1,s)$ and the symmetry
  \eqref{eq:sym_Q} of $Q$, 
\begin{multline}
  \int_{\mathcal{V}_{- 1}}\mu_{\mathcal{B}}(\d\nu'\d s')\tilde{\omega}(s')\Big[ Q((\nu',s'),(1, s))  + Q((\nu',s'),(-1, s)) - \delta(s'-s) + 2\int_{\mathcal{V}_0}\d \nu''\d s'' Q((\nu',s'),(0, s'')) Q((0,s''),(-1, s))\Big] \\=0.
\end{multline}
But the condition for $\nu=0$ is slightly different,
\begin{equation}
  2\int_{\mathcal{V}_{\pm
      1}}\d\mu_{\mathcal{B}}(\nu',s')\tilde{\omega}(s')Q((\nu',s'),(0,
  s)) \!=\! \tilde{\omega}(s)\mu_{\mathcal{B}}(0,s).
  \label{eq:muj0-decomp}
\end{equation} By unicity of
$\mu_{\mathcal{B}}$ defined by \eqref{eq:mub_cond}, we conclude,
\begin{equation}
\left\{  \begin{array}{l}
  \mu_{\mathcal{J}}(- 1,s)  =   2\mu_{\mathcal{B}}(\pm 1,s)\\
           \mu_{\mathcal{J}}(0,s) = \mu_{\mathcal{B}}(0, s)
           \end{array}\right. .
  \label{eq:muj-det}
  \end{equation}
  Therefore, in the case $K=Q$, the invariant measure $\pi$ is
  completely determined by the expression of
  $\mu_{\mathcal{B}}$. Furthermore, the detailed-jamming condition
  appears restrictive on the choices of $Q$, as
  \eqref{eq:wj-wb-cond} can be rewritten into this condition,
  \begin{equation}
\int_{\mathcal{V}_{1}}\d\mu_{\mathcal{B}}(1,s')\tilde{\omega}(s')(Q((1',s'),(-1, s))-Q((1,s'),(1, s)))=  \Big(\frac{2\phi(s)}{L}\frac{w_{\mathcal{B}}}{w_{\mathcal{J}}} -\omega(s)\Big)\mu_{\mathcal{B}}(1,s).
    \label{eq:wj-wb-cond-Q}
  \end{equation}

      \subsection{Global-jamming class}
      The detailed symmetry (\ref{eq:DB}) is not satisfied. This is
      for instance the case when the particles unjam at the smallest
      possible velocity while they can jam at biggest ones, implying
      some energy dissipation. Thus this class gathers systems further
      away from equilibrium, as we now quantify.

      First, while the active global balance (\ref{eq:all_flow}) still
      holds, there is no active detailed balance (\ref{eq:det_flow})
      at any point $r>0$ but at the periodic boundary $r=L/2$.  Sign
      of a stronger nonequilibrium nature, the stationary distribution
      $\pi_{\text{glob}}(r,\nu,s)$ can no longer be put under the form
      of Dirac contributions in $r=0$ and a product form in the bulk
      $r>0$, and is of the form,
   \begin{equation}
      \pi_{\text{glob}}(r,\nu,s)=  w_{\mathcal{J}}\1_{\{0\}}(r)\mu_{\mathcal{J}}(\nu,s)
        + w_{\mathcal{B}}\1_{]0,L/2[}(r)\mu(r,\nu,s),
     \label{eq:global_pi}
     \end{equation}
     where
     $$\mu(r,\nu,s)\!= \!w_{\text{eq}}\tfrac{2}{L}\mu_{\mathcal{B}}(\nu,s)\! +\!
     w_{\text{rel}}\gamma(r,\nu,s)$$ and $\gamma(\cdot)$ stands for
     the relaxation in the bulk from the non-detailed constraint on
     its jamming boundary to the detailed constraint on its periodic
     one. When plugging $\pi_{\text{glob}}$ into the bulk condition
     $(C_{\mathcal{B}})$ and exploiting the symmetry \eqref{eq:sym_Q}
     of $Q$, we obtain that the distributions $(\gamma(r,0,s))_s$ are determined by the
     sum of $\gamma(r,\pm 1, \cdot)$,
\begin{equation}
  \gamma(r, 0,  s)\tilde{\omega}(s) =   \int_{\mathcal{V}_1\cup\mathcal{V}_{-1}}\d s'\d \nu' \gamma(r,\nu,s')\tilde{\omega}(s') Q((1,s'), (0, s)), 
\end{equation}
and the distributions $\gamma(r,\pm 1, s)$ obey the relaxation equation,
  \begin{multline}
\pm \phi(s)\partial_r\gamma(r,\pm 1, s) \!=\! -   \gamma(r,\pm 1, s)\tilde{\omega}(s) 
+\int_{\mathcal{V}_1\cup\mathcal{V}_{-1}}\d s'\d \nu'  
\gamma(r,\nu, s')\tilde{\omega}(s') \\\qquad\times
\Big[Q((\nu,s'),(\pm 1, s)) 
\!+\!  \int_{\mathcal{V}_0}\d s''\d \nu''Q((1,s'), (0, s''))Q((0,s''),( 1, s)) \Big]
\label{eq:relax_continuous}
\end{multline}
Supposing that the internal states are of discrete values and noting
$\bm{\gamma}_\pm (r) = (\gamma(r,\pm 1, s))_{(1,s)\in\mathcal{V}_1}$ and
$\bm{\gamma}(r)=(\bm{\gamma}_+(r),\bm{\gamma}_-(r))$, the relaxation
equation \eqref{eq:relax_continuous} comes down to,
     \begin{equation}
         \partial_r\bm{\gamma}(r) = B\bm{\gamma}(r),
       \label{eq:edo}
     \end{equation}    
       where, by the symmetry
       (\ref{eq:sym_Q}), the bulk matrix $B$ is of the form
             $
         B = 
  \left( \begin{array}{cc}
           B^+ &   B^-\\
           -B^- &  -B^+
        \end{array}\right)$,
        with,
\begin{equation*}B^{+} = D^{-1}_{\phi}(Q^{+} - I)D_{\omega}, \ \ B^{-} = D^{-1}_{\phi}Q^{-}D_{\omega}  \in M_{|\mathcal{V}_1|\times|\mathcal{V}_1|}(\mathbb{R})
\end{equation*}
\begin{equation*}D_{\phi, ss'} = \phi(s)\delta_{ss'} \ \text{and} \ D_{\omega, ss'} = \omega(s)\delta_{ss'},
\end{equation*}
and,
\begin{equation*}Q^{\pm}_{ss'} = Q((\pm 1, s'), (1, \d s)) + \sum_{s''} Q((\pm 1, s'), (0, \d s'')) Q((0, s''), (1, \d s)).
\end{equation*}
As $Q$ is a Markov kernel and as there is no tumble between two
states with $\nu=0$ since it would require a tumble of each particle
at the same time, we have for any $(1,s')\in \mathcal{V}_1$ the conservation identity,
\begin{equation*} \sum_{s}(Q^{+}_{ss'} + Q^{-}_{ss'} - \delta_{ss'}) = 0
\end{equation*}
i.e.,
\begin{equation}
\sum_{s}\phi(s)\big(B^+ +  B^{-}\big)_{ss'}=0,
\label{eq:tr_0}
\end{equation}    
which is consistent with the active global balance condition
(\ref{eq:all_flow}), as clearly appearing when rewritten as,
\begin{equation*}
\sum_s\phi(s)(\partial_r\bm{\gamma}_s^+(r)-\partial_r\bm{\gamma}_s^-(r)) = 0 =  \sum_{s}\phi(s)\big(B^+ +  B^{-}\big)_{ss'}(\bm{\gamma}_{s'}^+(r)+\bm{\gamma}_{s'}^-(r)).
\end{equation*}

The symmetry (\ref{eq:sym_Q}) also constrains the Jordan form of $B$,
as it admits as eigenvalues symmetrical finite $(\pm \lambda_k)_k$ of
respective eigenvectors $(\bm{\mu}^\pm_{k,1},\bm{\mu}^\mp_{k,1})$ and
necessarily the eigenvalue $0$ of eigenvector
$(\bm{\mu}_{\mathcal{B}},\bm{\mu}_{\mathcal{B}})$, i.e. corresponding
to the stationary state of the tumble process with
$\bm{\mu}_{\mathcal{B}}=(\mu_{\mathcal{B}}(1,
s))_{(1,s)\in\mathcal{V}_1}$. A longer discussion and justifications
can be found in \ref{app:spectrum}.

In particular, the spectral properties of $B$ are strongly related to
the one of $M^+M^-$ and $M^-M^+$, with $M^\pm=B^+\pm B^-$. The
characteristic and minimal polynomials of $B$ are indeed written in
terms of $B^2$, which is made of blocks $M^-M^+ \pm M^+M^-$. The
generalized eigenvectors of $M^- M^+$ (resp. of $M^+M^-$) identify
with
$((\bm{\mu}^+_{k,n}+\bm{\mu}^-_{k,n})_{k,1<n\leq p_k},
(\bm{\mu}_{0,2n-1})_{1<n\leq p_0/2})$ (resp.
$((\bm{\mu}^+_{k,n}-\bm{\mu}^-_{k,n})_{k,1<n\leq p_k},
(\bm{\mu}_{0,2n})_{1< n\leq p_0/2})$).

Therefore, studying the squared global bulk matrix $B^2$ or
equivalently the submatrices $M^+M^-$ and $M^-M^+$, seems the natural
way to deal with the bulk evolution. Indeed, it underlines the fact
that, instead of considering the global bulk evolution
$\partial_r\bm{\gamma} = B\bm{\gamma}$, one can consider the system
composed of the symmetrical bulk evolution,
\begin{equation}
\partial_r(\bm{\gamma}_+ + \bm{\gamma}_-) = M^-(\bm{\gamma}_+ - \bm{\gamma}_-),
\label{eq:sym_bulk}
\end{equation}
and  of the antisymmetrical bulk evolution,
\begin{equation}
\partial_r(\bm{\gamma}_+ - \bm{\gamma}_-) = M^+(\bm{\gamma}_+ + \bm{\gamma}_-).
\label{eq:sym_bulk}
\end{equation}
It gives the uncoupled second-order system,
\begin{equation}
\left\{  \begin{array}{l}
           \partial^2_r(\bm{\gamma}_+ \!+\! \bm{\gamma}_-) = M^-M^+(\bm{\gamma}_+\! +\! \bm{\gamma}_-)\\
           \partial^2_r(\bm{\gamma}_+ \!-\! \bm{\gamma}_-) = M^+M^-(\bm{\gamma}_+ \!-\! \bm{\gamma}_-)
           \end{array}\right.,
  \label{eq:main_second_order}
\end{equation}
which leads to catenary-like solutions in common cases, as illustrated
below and in Section~\ref{sec:app-3rtp-gen}. 

As detailed in \ref{app:gen_bulk}, the general solution of
(\ref{eq:edo}) is
\begin{equation}
  \begin{aligned}
    \bm{\gamma}^+(r) &= \bm{\gamma}^{L}|_{[0,L/2]}(r)  \text{ and }  \bm{\gamma}^-(r) = \bm{\gamma}^{L}|_{[L/2,L]}(L-r), \text{ with } \bm{\gamma}^L:[0,L] \to \mathbb{R} \text{ as,}\\
    \bm{\gamma}^{L}(r)&= \sum_{l=0}^{p_0/2-1}b_{0,2l+1} \frac{\big(\tfrac{L}{2}-r\big)^{2l}}{(2l)!}\bm{\mu}_{\mathcal{B}}  \\&
 \displaystyle\,\,\,\,\,+
\sum_{n=1}^{p_0/2-1}\sum_{l=n}^{p_0/2-1}b_{0,2l+1} \frac{\big(\tfrac{L}{2}-r\big)^{2(l-n)}}{(2(l-n))!}\Big[
\bm{\mu}_{0,2n+1} 
-\frac{\big(\tfrac{L}{2}-r\big)}{2(l-n)+1}
\bm{\mu}_{0,2n}  \Big]\\  
& \displaystyle \,\,\,\,\,  + \sum_{k=1}^{d}\sum_{n=1}^{p_k} \sum_{l=n}^{p_k}b_{k,l}\frac{\big(\tfrac{L}{2}-r\big)^{l-n}}{(l-n)!} \Big[ (-1)^{l-n} e^{-\lambda_k\left(\frac{L}{2}-r\right)}
\bm{\mu}_{k,n}^+
+ e^{\lambda_k\left(\frac{L}{2}-r\right)}
\bm{\mu}_{k,n}^-
\Big] . 
\end{aligned}
\label{eq:sol_main_expl}  
\end{equation} 
It derives from the spectral properties of $B$ and decomposes over a
sum of exponentials of rate $(\pm \lambda_k)_k$, modulated by
polynomial terms stemming from possible degeneracies of the
$\lambda_k$, as detailed in \ref{app:bulk}. The symmetry
$\bm{\gamma}^+(r) = \bm{\gamma}^-(L-r)$ (formally,
  when defined on the whole torus $[0,L]$) arises from the periodic
boundary constraint, as discussed in \ref{app:per-cond}, and the
precise decomposition is set by the boundary conditions at jamming. As
$B$ admits $0$ as an eigenvalue, the general solution also admits a
pure polynomial term.

An interesting case is when $B$ does not admit Jordan blocks
of size bigger than $3$, as, from $(C_{P})$, any polynomial decay
disappears and we recover an exact symmetry between decaying and
increasing exponentials and such for any jamming scenario, leading to,
    \begin{equation}
            \bm{\gamma}_\pm(r)\!=\!                            
     \sum_{k} a_k\big[ \cosh\Big(\lambda_k\Big(\frac{L}{2}\!-\!r\Big)\Big) 
          (\bm{\mu}_{k,1}^+  \!+\! \bm{\mu}_{k,1}^-) 
    \!\mp \sinh\Big(\lambda_k\Big(\frac{L}{2}\!-\!r\Big)\Big)
       ( \bm{\mu}_{k,1}^+   \!-\!\bm{\mu}_{k,1}^-)\big],   
\label{eq:main_cat_sol}  
\end{equation}
with $(a_k)_k$ set by $(C_{\mathcal{J}})$. It yields a catenary-like
relaxation once marginalized over $(\nu,s)$, as hinted by
(\ref{eq:main_second_order}). A particular case is any system with
particle internal states symmetrical enough so that $B$ only admits a
single non-null $\pm \lambda_k$, so that
$\lambda_k^2\propto\text{Tr}((B^++B^-)(B^+-B^-))$, as for the
finite tumble (see Section~\ref{sec:app-3rtp}), which simplifies the
formula in \cite{slowman2017}.\\

{\noindent \bf Case $K=Q$.} Therefore, the richness of the relaxation
behavior directly arises from the inherent spectral richness of the
bulk matrix itself, i.e. the tumbling mechanism, to the extent of what
is unveiled by the unjamming scenario. In \ref{app:jam_cond}, we give
more details on the jamming boundary constraint in the case $K=Q$. Notably, we note $\bm{\mu}_{\mathcal{J}}=(\mu_{\mathcal{J}}(-1,s))_{(-1,s)\in\mathcal{V}_{-1}}$ and show that
\begin{equation}    
\left\{
\begin{array}{l}
w_{\mathcal{J}}M^+\bm{\mu}_{\mathcal{J}}
=  w_{\mathcal{B}}w_{\text{rel}}(\bm{\gamma}^+(0^+) -\bm{\gamma}^-(0^+))\\      
-w_{\mathcal{J}}M^-\bm{\mu}_{\mathcal{J}}
=  w_{\mathcal{B}} (2w_{\text{eq}}\bm{\mu}_{\mathcal{B}} + w_{\text{rel}}(\bm{\gamma}^+(0^+) +\bm{\gamma}^-(0^+)))
\end{array}
\right ..
\label{eq:0_plus-main}
\end{equation}
It clearly appears that recovering a detailed-jamming solution
(i.e. $w_{\text{rel}}=0$) imposes $M^+\bm{\mu}_{\mathcal{J}}=0$,
i.e. $\bm{\mu}_{\mathcal{J}}\propto\bm{\mu}_{\mathcal{B}}$, which,
combined with the second condition, then makes the detailed-jamming
condition equivalent to $\bm{\mu}_{\mathcal{B}}$ being the unique
eigenvector shared by $B^+$ and $B^-$ with opposite eigenvalues. It
also imposes that
$M^-\bm{\mu}_{\mathcal{J}}\propto\bm{\mu}_{\mathcal{B}}$, i.e.
$\bm{\mu}_{\mathcal{J}}\propto\bm{\mu}_{0,2}$, which leads to the
condition $\bm{\mu}_{0,2}\propto \bm{\mu}_{\mathcal{B}}$, which is
consistent with the condition \eqref{eq:wj-wb-cond-Q}.  In the global-jamming case,
\eqref{eq:0_plus-main} sets the value of the different weights, once
injecting the decompositon of $\bm{\mu}_{\mathcal{J}}$ along the basis
of generalized eigenvectors.

\medskip
In the following Sections~\ref{sec:app-2rtp} and \ref{sec:app-3rtp-gen}, we give
explicit subclasses which illustrate behaviors of interest as
detailed-jamming symmetry, pure catenary-like relaxation and
multiscale catenary-like relaxation.

\section{Subclasses of interest: General two-state RTP systems as a subclass of the detailed-jamming class}
\label{sec:app-2rtp}

An interesting subclass in the detailed-jamming class is systems with
two internal states $v \in \{v_1,v_2\}$. This includes in particular
the case of the symmetrical instantaneous tumble ($v_1,v_2=-1,+1$,
$\omega(v_1)=\omega(v_2)$), as illustrated in
Figure~\ref{fig:transition_rates}. These systems are always
detailed-jamming ones, as there is only one entry state into jamming
and one exit state out of jamming, realizing the detailed-jamming
symmetry (\ref{eq:DB}). For these systems and considering the case
$K=Q$, we obtain,
\[ 
\def\arraystretch{1.4}
\left\{
\begin{array}{l}
\omega(v_2)\mu_{\mathcal{B}}(\pm 1, \{v_1,v_2\})
=   \omega(v_1)\mu_{\mathcal{B}}(0, \{v_1,v_1\})\\
\omega(v_1)\mu_{\mathcal{B}}(\pm 1, \{v_1,v_2\})
=   \omega(v_2)\mu_{\mathcal{B}}(0, \{v_2,v_2\})\\
\end{array}\right..\]  
We then get, by \eqref{eq:muj-det},
\begin{equation}
\begin{array}{llll}
\def\arraystretch{1.4}
\left\{      \begin{array}{l}
\mu_{\mathcal{J}}(0,\{v_1,v_1\}) = \tfrac{\omega(v_2)^2}{\mathcal{N}_{\mathcal{J}}}\\
\mu_{\mathcal{J}}(0,\{v_2,v_2\}) = \tfrac{\omega(v_1)^2}{\mathcal{N}_{\mathcal{J}}}\\
\mu_{\mathcal{J}}(-1,\{v_1,v_2\}) = 2 \tfrac{\omega(v_1)\omega(v_2)}{\mathcal{N}_{\mathcal{J}}}\\
\mathcal{N}_{\mathcal{J}}=\big(\omega(v_1)+\omega(v_2)\big)^2
\end{array}\right.,
&  & \def\arraystretch{1.4}\left\{\begin{array}{l}
\mu_{\mathcal{B}}(0,\{v_1,v_1\}) = \tfrac{\omega(v_2)^2}{\mathcal{N}_{\mathcal{B}}}\\
\mu_{\mathcal{B}}(0,\{v_2,v_2\}) = \tfrac{\omega(v_1)^2}{\mathcal{N}_{\mathcal{B}}}\\
\mu_{\mathcal{B}}(\pm 1,\{v_1,v_2\}) =  \tfrac{\omega(v_1)\omega(v_2)}{\mathcal{N}_{\mathcal{B}}}\\
\mathcal{N}_{\mathcal{B}}=\big(\omega(v_1)+\omega(v_2)\big)^2
\end{array}\right.,
\end{array}
\label{eq:mub-muj-2state}
\end{equation}
and by \eqref{eq:wj-wb-cond},
\begin{equation*}
\left\{\begin{array}{l}
w_{\mathcal{B}} =\frac{L(\omega(v_1)+\omega(v_2))}{L(\omega(v_1)+\omega(v_2))+ 2|v_1-v_2|}\\

w_{\mathcal{J}}/ w_{\mathcal{B}}= \frac{2|v_1-v_2|}{L(\omega(v_1)+\omega(v_2))}
\end{array}\right. .
\end{equation*}
Therefore the ratio $w_{\mathcal{J}}/ w_{\mathcal{B}}$ of time spent
in jamming versus in the bulk increases with the strength of the
ballistic nature of the considered RTP dynamics (decrease of the
time needed to cover the length $L$ of the torus, decrease of the
total tumble rate).

\section{Subclasses of interest: General three-state RTP systems as
  subclass of the global-jamming class}
\label{sec:app-3rtp-gen}

In this section, we consider RTP with a three-state dynamical space
$\{v_1,v_2,v_3\}$ and the case $K=Q$. Based on \eqref{eq:wj-wb-cond-Q},
such systems can never belong to the detailed-jamming class. In the
following, we cover systems with a pure catenary relaxation, thanks to
some additional symmetry, and systems with multiscale catenary
relaxation.

\subsection{Symmetrical three-state RTP systems and single non-null
  eigenvalue relaxation}
\label{sec:app-3rtp}

An interesting situation is when the particles have an internal state
space and tumbling mechanism presenting enough symmetries so that it
enables a collapsed representation, through the choice of the $s$ variable,
where $\mathcal{V}_{\pm 1}$ is a two-state space. In this collapsed
representation, the corresponding matrix $B_{\text{coll}}$ is of size
$4$. As the Jordan block associated to $0$ is at least of size $2$,
$B_{\text{coll}}$ can only admit two Jordan blocks of size $1$ linked
to the eigenvalues $\pm\lambda\neq 0$. From the spectral analysis of
$B_{\text{coll}}$, as detailed in \ref{app:submat}, $\lambda^2$ is the
only non-null eigenvalue of $M_{\text{coll}}^+M_{\text{coll}}^-$ (or
$M_{\text{coll}}^-M_{\text{coll}}^+$), so that
$$\lambda^2=\text{Tr}(M_{\text{coll}}^+M_{\text{coll}}^-).$$
Therefore, in the bulk, the invariant measure $\pi$ follows a
single-scale pure catenary relaxation \eqref{eq:main_cat_sol} set by
$\lambda$.

\begin{figure}[!h]
	\centering
	\includegraphics[width=.6\textwidth]{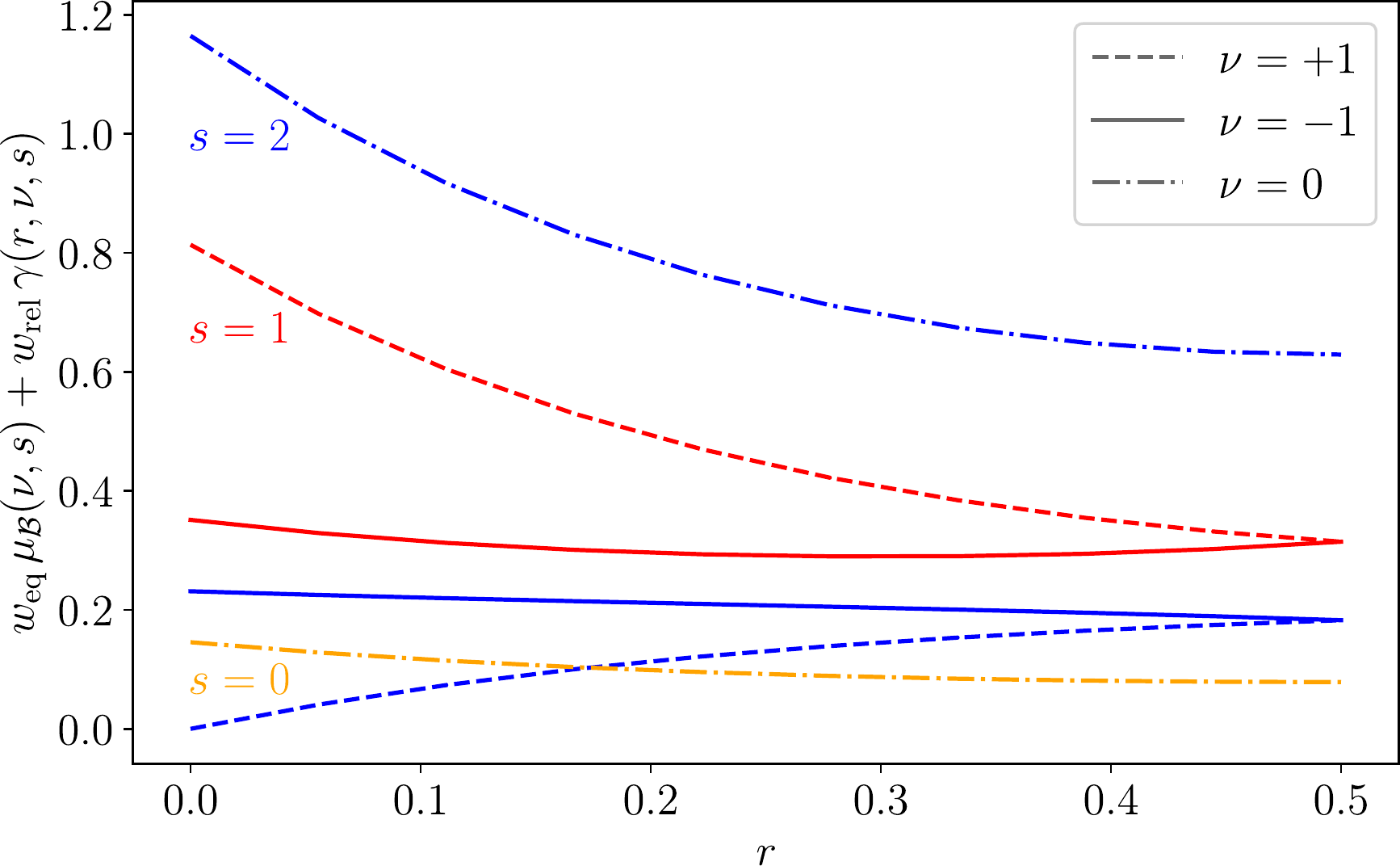}
	\caption{Density of the invariant measure in the bulk for $\alpha=1$, $\beta=4$ and $L=1$}
	\label{fig:finite_tumble_densities}
\end{figure}

This is the case of
the finite tumble represented in Figure~\ref{fig:transition_rates} we
now use as an illustrative example. For such systems, the standard representation
with the choice $s=\{v_1,v_2\}$ can be collapsed to the one with
$s_{\text{coll},t} =
\mathbb{1}_{\{-1,1\}}(v_{t,1})+\mathbb{1}_{\{-1,1\}}(v_{t,2})$, which
counts the number of moving particles. The typical inverse length scale of the
catenary relaxation is,
\begin{equation*}
\lambda = \sqrt{(\alpha +\beta)(\alpha+\tfrac{1}{2}\beta)} 
\ \text{and} \
  \bm{\mu}_{\mathcal{B}}=(\mu_{\mathcal{B}}(\pm 1, 1),\mu_{\mathcal{B}}(\pm 1, 2)) \propto (4\beta,\alpha).    
\end{equation*}
The density $\pi$ of the invariant measure in the bulk is represented
in Figure~\ref{fig:finite_tumble_densities} in the finite tumble
case.
In particular, Figure~\ref{fig:finite_tumble_densities} illustrates
the unbalance between $\pi(0^+, 1, s) \neq \pi(0^+,- 1, s) $ at the
jamming boundary and the catenary relaxation towards the detailed
symmetry $\pi(L/2, 1, s) = \pi(L/2,- 1, s) $ at the periodic boundary.

\begin{figure}[H]
	\centering
	\begin{subfigure}[b]{0.32\textwidth}
		\centering
		\includegraphics[width=\textwidth]{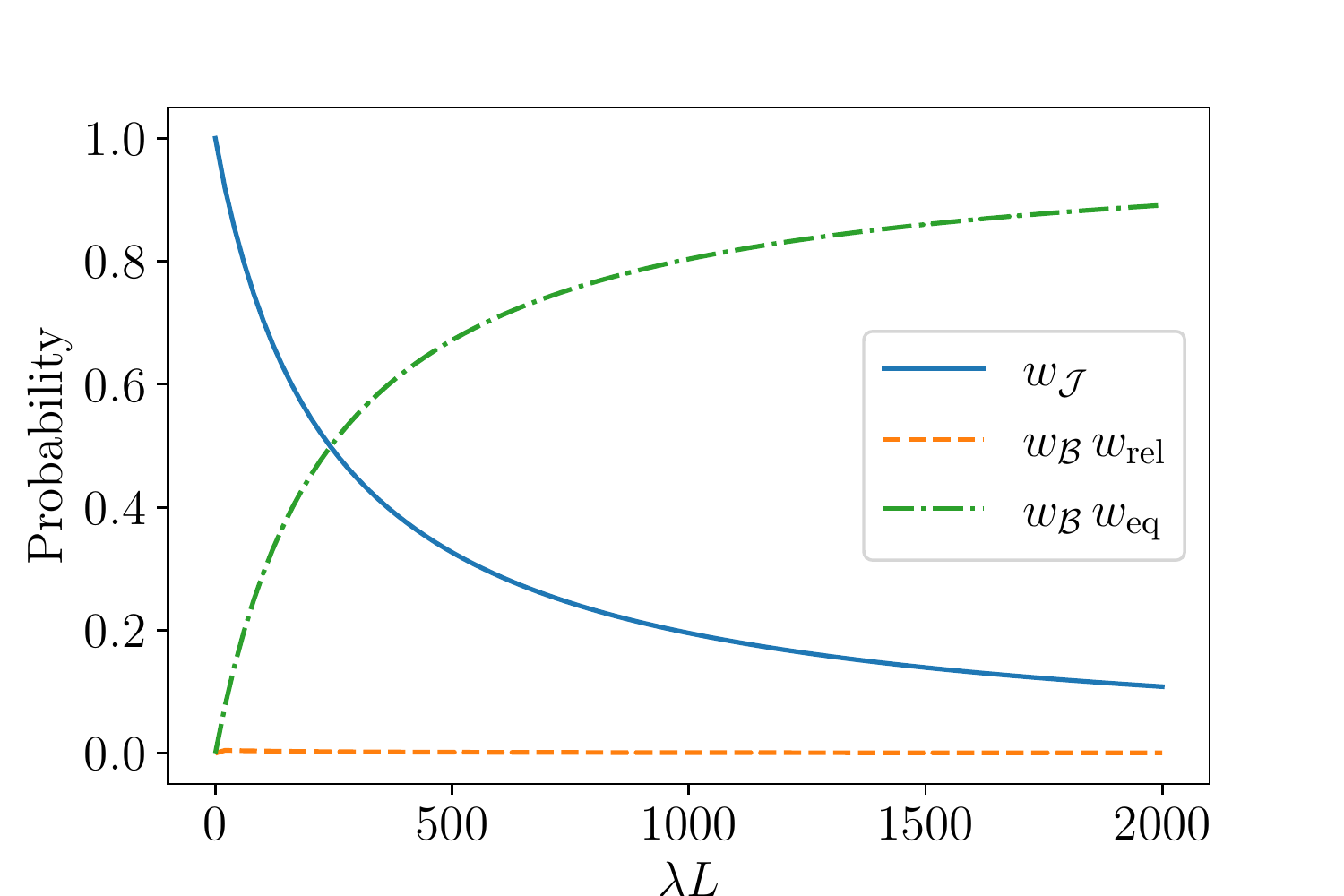}
		\caption{$\frac{\alpha}{\beta}=0.01$}
	\label{fig:rel_weigths_finite_tumble_r=.01}
	\end{subfigure}
	\hfill
	\begin{subfigure}[b]{0.32\textwidth}
		\centering
		\includegraphics[width=\textwidth]{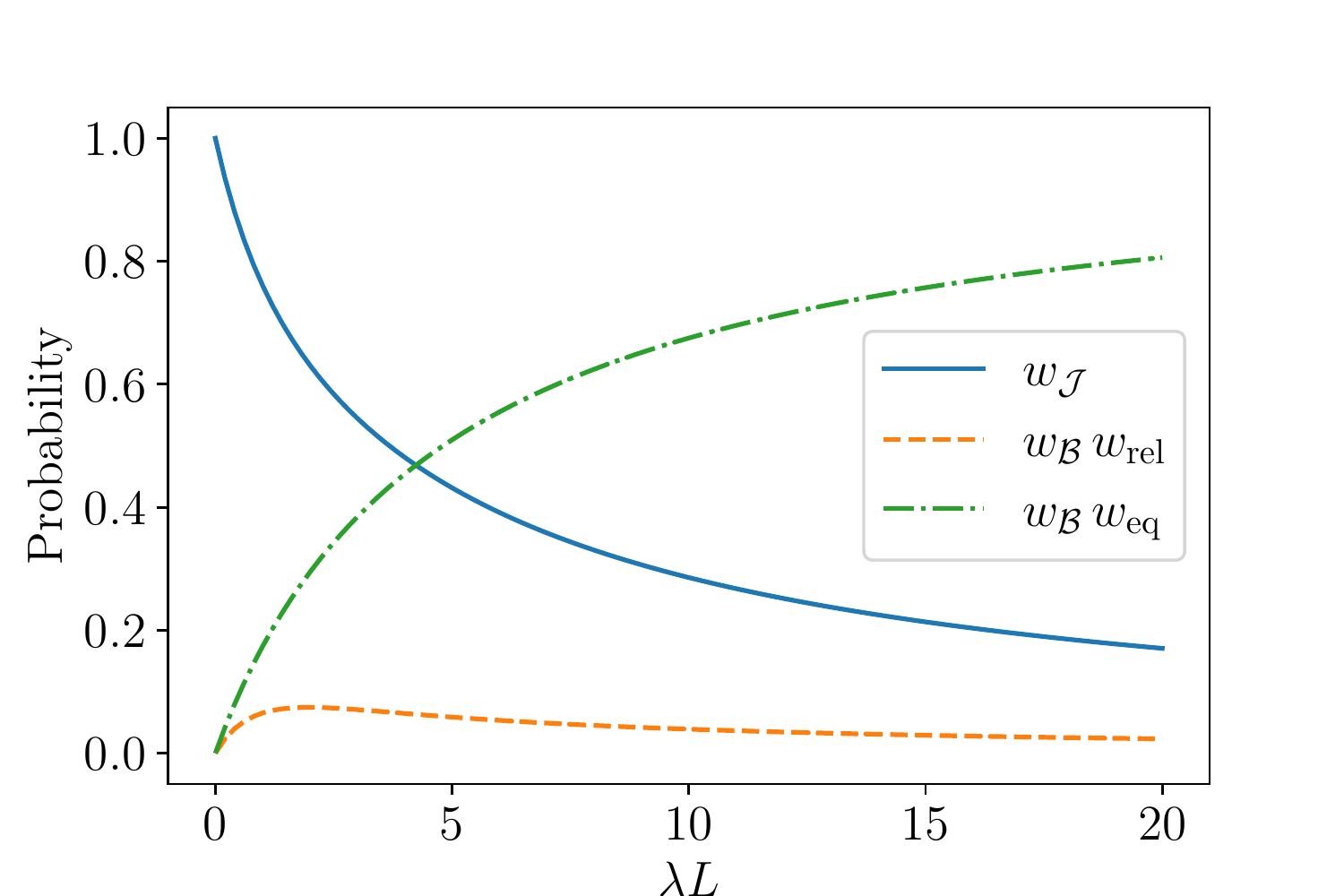}
		\caption{$\frac{\alpha}{\beta}=1$}
	\label{fig:rel_weigths_finite_tumble_r=1}
	\end{subfigure}
	\hfill
	\begin{subfigure}[b]{0.32\textwidth}
		\centering
		\includegraphics[width=\textwidth]{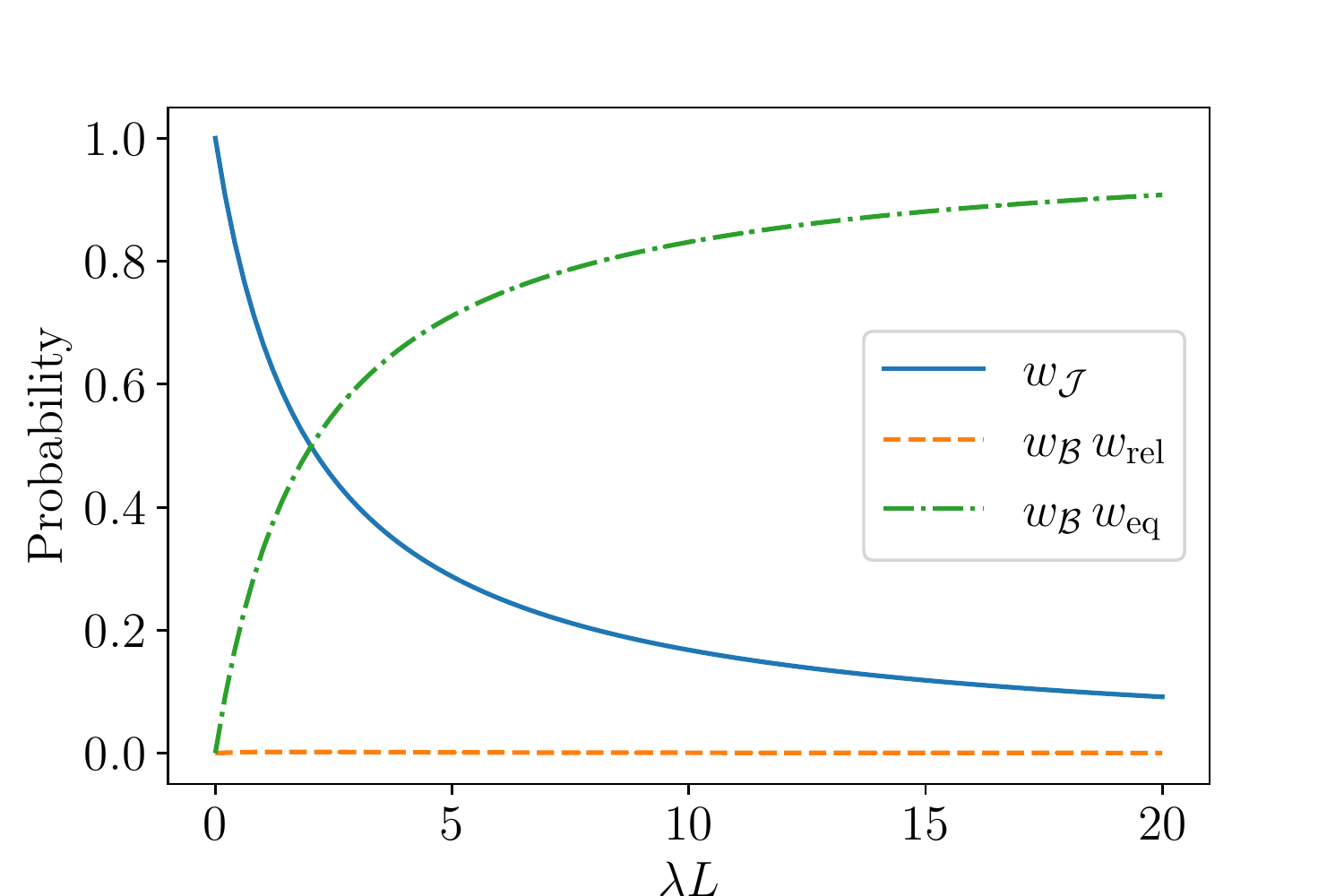}
		\caption{$\frac{\alpha}{\beta}=100$}
	\label{fig:rel_weigths_finite_tumble_r=100}
	\end{subfigure}
\caption{Respective contributions of the Dirac, exponential and uniform terms towards the invariant measure}
	\label{fig:rel_weigths_finite_tumble}
\end{figure}

The evolution along $\lambda L$ of the respective contributions of the
Dirac masses ($w_\mathcal{J}$), relaxation
($w_{\mathcal{B}}w_{\text{rel}}$) and uniform terms
($w_{\mathcal{B}}w_{\text{rel}}$) are plotted on
Figure~\ref{fig:rel_weigths_finite_tumble} for different ratio
$\alpha/\beta$.  Once the velocity and ratio $\alpha / \beta$ are
fixed, these contributions indeed only depend on the parameter
$\lambda L$. Such parameter can be thought of of as a measure of
ballisticity/diffusivity. Indeed, when $\lambda L \ll 1$ the average
time between two tumbles (order $1/\lambda$) is much larger than the
time needed to reach a jammed state in the absence of velocity flips
(order $L$) so the particle movement is overwhelmingly ballistic and
the system does not decorrelate from the jamming constraint. On the
other hand, when $\lambda L \gg 1$ the particles constantly change
direction, their movement pattern is well approximated by a diffusion
and, in the bulk, the density $\pi$ immediately relaxes to the
uniform solution set by $\tfrac{2}{L}\mu_{\mathcal{B}}$.

\begin{figure}
 \includegraphics[width=0.8\textwidth]{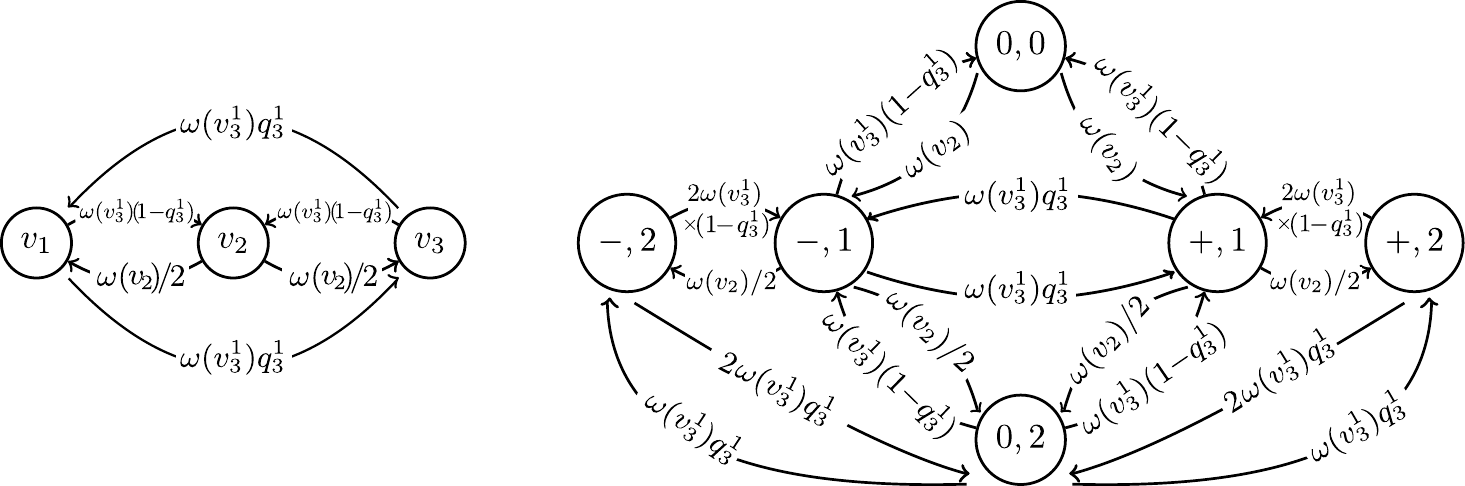}
 \caption{Transition rates (i.e. $\omega(\cdot)q(\cdot,\cdot)$)
   between the internal states of a single symmetrical 3-state RTP
   ({\bf left}) and corresponding transition rates
   (i.e. $\tilde{\omega}(\cdot)Q(\cdot,\cdot)$) at the interdistance
   level ({\bf bottom}) in the representation
   $s_t = \mathbb{1}_{\{v_1,v_3\}}(v_{t,1}) +
   \mathbb{1}_{\{v_1,v_3\}}(v_{t,2})$ ({\bf right}). The notations
   $\omega(v^1_3)=\omega(v_1)=\omega(v_3)$ and
   $q^1_3= q(v_1,v_3)=1(v_3,v_1)$ are used.}
\label{fig:3-state}
\end{figure}

We now consider the general case of the symmetrical three-state RTP system, as
presented in Figure~\ref{fig:3-state}. This is a generalization of the
finite tumble case of Figure \ref{fig:transition_rates}. We note the three states
$\{v_1,v_2,v_3\}$. The symmetry leading to a possible collapse to a
four-state space is obeyed if two pair states, e.g. $\{v_1,v_2\}$ and
$\{v_2,v_3\}$ are equivalent, i.e. present the same run and tumble
mechanisms, leading to
$\pi(r,\pm 1,\{v_2,v_3\})=\pi(r,\pm 1,\{v_1,v_2\})$, and more
precisely framed by the conditions,
\begin{equation*}
\left\{\begin{array}{l}
q(v_3,v_1)=q(v_1,v_3)=q_{1,3}\\
q(v_2,v_1)=q(v_2,v_3)= \tfrac{1}{2} \\
q(v_1,v_2)=q(v_3,v_2)=1-q_{1,3}\\
|v_3-v_2|=|v_2-v_1|=|v_{1,3}-v_2|\\
\omega(v_3) = \omega(v_1) = \omega_{1,3}\\
\omega(v_2) = \omega_2\\
\end{array}\right..
\end{equation*}
Note that, by the ergodicity constraint on the tumble mechanism,
$0\leq q_{1,3}<1$. The finite tumble corresponds to
$q(v_1,v_3)=q(v_3,v_1)=0$, $v_1=-1, v_2=0, v_3=1$ and
$\omega(v_1)=\omega(v_3) = \alpha$, $\omega(v_2) = \beta$.

We obtain the following matrices,
 $$M_{\text{coll}}^+ = \bordermatrix{ &  s = 1 & s = 2   \cr
	s = 1 & -\omega_2/2 &  2\omega_{1,3}(1-q_{1,3})  \cr
		      &                  &                       \cr
	s = 2 & \omega_2/4 & -\omega_{1,3}(1-q_{1,3}) }\frac{(1+q_{1,3})}{|v_{1,3}-v_2|} 
      $$
 and,
$$M^-_{\text{coll}} = \bordermatrix{ &  s = 1 & s = 2   \cr
	s = 1 & -\omega_{1,3}(1+q_{1,3})-\omega_2   & 2\omega_{1,3}(1-q_{1,3})  \cr
		      &                  &                       \cr
	s = 2 &\omega_2/4 & -\omega_{1,3} } \frac{1}{|v_{1,3}-v_2|}
      $$  
that are collapsed variants of $M^+$ and $M^-$.
 $$M^+ = \bordermatrix{ &  s = \{v_1,v_2\} & s = \{v_2,v_3\} & s=\{v_3,v_1\}   \cr
	s = \{v_1,v_2\} & -(\omega_{1,3}+\omega_2)/2 & \omega_{1,3}/2 & \omega_{1,3}( 1-q_{1,3}) \cr
		      &      &            &                       \cr
	s = \{v_2,v_3\} &  \omega_{1,3}/2 &-(\omega_{1,3}+\omega_2)/2 & \omega_{1,3}( 1-q_{1,3}) \cr 
		      &     &             &                      \cr
	s = \{v_3,v_1\} & \omega_2/4 &  \omega_2/4 &  -\omega_{1,3}(1-q_{1,3}) \cr }\frac{(1+q_{1,3})}{|v_{1,3}-v_2|} 
      $$
 and,
 $$M^- = \bordermatrix{ &  s = \{v_1,v_2\} & s = \{v_2,v_3\} & s=\{v_3,v_1\}   \cr
	s = \{v_1,v_2\} & -(\omega_{1,3}+\omega_2)  & -\omega_{1,3}q_{1,3} &\omega_{1,3}(1-q_{1,3})  \cr
		      &      &            &                       \cr
	s = \{v_2,v_3\} & -\omega_{1,3}q_{1,3}& -(\omega_{1,3}+\omega_2) & \omega_{1,3}(1-q_{1,3})\cr 
		      &     &             &                       \cr
	s = \{v_3,v_1\} &  \omega_2/4&\omega_2/4 &-\omega_{1,3} \cr }\frac{1}{|v_{1,3}-v_2|}.
      $$
Then, the single non-null eigenvalue $\lambda$ is such that,
\begin{equation}
\lambda^2\! =\! \Tr(M^+_{\text{coll}}M^-_{\text{coll}}) =(\omega_{1,3} (1-q_{1,3})+\omega_2)(\omega_{1,3}+\tfrac{1}{2}\omega_2) \frac{1+q_{1,3}}{(v_{1,3}-v_2)^2}.
\end{equation}
It is noteworthy that $\lambda^2 > 0$ for any parameter values, as
expected given that the bulk stationary distribution, corresponding to
the eigenvector
$ \bm{\mu}_{\mathcal{B}}=(\mu_{\mathcal{B}}(\pm
1,1),\mu_{\mathcal{B}}(\pm 1, 2))$ of eigenvalue $0$, is
unique. However, it may still be possible to recover a
detailed-jamming solution, as the coefficients $(a_k)_k$ are set by
$(C_{\mathcal{J}})$ in \eqref{eq:main_cat_sol}. As further explained
in \ref{app:jam_cond}, the condition $(C_{\mathcal{J}})$ may yield a
detailed-jamming condition (i.e. $a_k = 0, \forall k$) if and only if
the generalized eigenvector $\bm{\mu}_{0,2}$ is proportional to
$\bm{\mu}_{\mathcal{B}}$. Here, from the relationship
$M^+_{\text{coll}}\bm{\mu}_{\mathcal{B}}=0$ and
$M^-_{\text{coll}}\bm{\mu}_{0,2}=\bm{\mu}_{\mathcal{B}}$, we get,
\begin{equation}
\left\{\begin{array}{l}
  \bm{\mu}_{\mathcal{B}} \propto (4\omega_{1,3}(1-q_{1,3}),\omega_2)\\         
  \bm{\mu}_{0,2} \propto (2\omega_{1,3}(1-q_{1,3}), \omega_2)\\
       \end{array}\right.,
     \label{eq:mub_unnormed}
\end{equation}
so that a detailed-jamming condition would impose $q_{1,3}=1$, which
is not possible. It actually transforms the symmetrical 3-state RTP
system back to a detailed-jamming 2-state RTP one as discussed in the
previous section. Finally, $\lambda$ is the inverse relaxation length
scale of the system. Therefore $\lambda L \gg 1$ corresponds to
an instant tumble thermalization leading to a diffusive regime and
converging towards a detailed-jamming solution. On the other hand, the
regime $\lambda L << 1$ corresponds to a ballistic one. This is
confirmed by the following analysis.

Reminding first the general form of a
global-jamming stationary distribution \eqref{eq:global_pi},
$$ 
\pi_{\text{glob}}(r,\nu,s)=
w_{\mathcal{J}}\1_{\{0\}}(r)\mu_{\mathcal{J}}(\nu,s)
+w_{\mathcal{B}}\1_{]0,L/2[}(r)\big(w_{\text{rel}}\gamma(r,\nu,s) +
w_{\text{eq}}\tfrac{2}{L}\mu_{\mathcal{B}}(\nu,s)\big).
$$
From \eqref{eq:mub_unnormed} and $(C_{\mathcal{B}})$ (or
\eqref{eq:mub_cond}), we have for the tumble stationary distribution,
\begin{equation}
\left(\begin{array}{l}
        \mu_{\mathcal{B}}(\pm 1,1) \\
        \mu_{\mathcal{B}}(\pm 1,2)   \\
        \mu_{\mathcal{B}}(0, 0)        \\
        \mu_{\mathcal{B}}(0, 2)
      \end{array}\right)=\frac{1}{\mathcal{N}_{\mathcal{B}}}
   \left(    \begin{array}{l}
                4\omega_{1,3}\omega_2(1-q_{1,3})\\
                \omega_2^2\\
               4\omega_{1,3}^2(1-q_{1,3})^2 \\
                2\omega_2^2\\
             \end{array}\right),
           \label{eq:mub_normed}
\end{equation}
with,
$$
\mathcal{N}_{\mathcal{B}}=4(\omega_{1,3}(1-q_{1,3}) + \omega_2)^2.
$$
 In the limit $q_{1,3}\to 1$, the distribution expression correctly
identifies with the 2-state RTP system \eqref{eq:mub-muj-2state}. Therefore, this
fixes
\begin{equation}
  \bm{\mu}_{0,2}=\frac{-2|v_{1,3}-v_2|}{\mathcal{N}_{\mathcal{B}}}\frac{\omega_2}{\omega_{1,3}(1+q_{1,3}))}
\left( \begin{array}{l}2\omega_{1,3}(1-q_{1,3})\\
         \omega_2\end{array}\right).
     \label{eq:mu02}
     \end{equation}
We obtain as the relaxation term,
\begin{equation}
  \Big( \begin{array}{l}
          \gamma(r,\pm 1, 1) \\
          \gamma(r,\pm 1,2)
        \end{array}\Big)
        \!=\!                            
            \cosh\Big(\lambda\Big(\frac{L}{2}\!-\!r\Big)\Big) 
          (\bm{\mu}_{1}^+  \!+\! \bm{\mu}_{1}^-) 
    \!\mp  \sinh\Big(\lambda\Big(\frac{L}{2}\!-\!r\Big)\Big)
       ( \bm{\mu}_{1}^+   \!-\!\bm{\mu}_{1}^-),   
\label{eq:3-state-gamma}  
  \end{equation}
  with
  $\bm{\mu}_{1}^+ \! \pm \!\bm{\mu}_{1}^- $ eigenvectors of
  respectively $M_{\text{coll}}^\mp M_{\text{coll}}^\pm$ of eigenvalue $\lambda^2$, that are,
\begin{equation}
  \bm{\mu}_1^+ + \bm{\mu}_1^- =  \frac{1}{\mathcal{N}_\gamma}\begin{pmatrix} 4 \\ -1 \end{pmatrix} \text{ and } \bm{\mu}_1^+ - \bm{\mu}_1^- =  \frac{1}{\mathcal{N}_\gamma}\frac{2\lambda|v_{1,3}-v_2|}{2\omega_{1,3}+\omega_2}\begin{pmatrix} -2 \\ 1 \end{pmatrix},
  \label{eq:mu1pm}
\end{equation}
with 
  $$\mathcal{N}_\gamma=\frac{4\lambda\sinh\big(\lambda\tfrac{L}{2}\big)}{\omega_{1,3}\omega_2}\frac{(v_{1,3}-v_2)^2}{1+q_{1,3}},$$
  some normalization so that $\gamma(\cdot)$ is normalized.
Eventually, we get,
\begin{equation}
  \Big( \begin{array}{l}
          \gamma(r,0, 0) \\
          \gamma(r,0,2)
        \end{array}\Big)
        \!=\!                            
        \frac{1}{\mathcal{N}_\gamma}\frac{2\cosh\Big(\lambda\Big(\frac{L}{2}\!-\!r\Big)\Big)}{\omega_{1,3}\omega_2}\left(  \begin{array}{l}  2\omega_{1,3}^2(1-q_{1,3})
                                                                                                   \\
                                                                                                   \omega_2(\omega_2-\omega_{1,3}q_{1,3})
        \end{array}\right),   
\label{eq:3-state-gamma-0}  
  \end{equation}
  We now determine the weights
  $w_{\text{eq}}, w_{\text{rel}}, w_{\mathcal{J}},
  w_{\mathcal{B}}$. First, the value of the coefficient
  $w_{\text{eq}}=1-w_{\text{rel}}$ stems from the breaking of the
  detailed-jamming symmetry, leading to
  $\bm{\mu}_{0,2}\not\propto \bm{\mu}_{\mathcal{B}}$ and
  $\bm{\mu}_{\mathcal{J}}\not\propto
  \bm{\mu}_{\mathcal{B}}$. Therefore, such discrepancy can be
  determined thanks to the decomposition of
  $\bm{\mu}_{\mathcal{J}}=(\mu_{\mathcal{J}}(-1,1),\mu_{\mathcal{J}}(-1,2))$
  over the two following basis, as discussed in \ref{app:jam_cond},
  \begin{align}
    \bm{\mu}_{\mathcal{J}}&=-2\frac{w_{\mathcal{B}}}{w_{\mathcal{J}}}\Big(w_{\text{eq}}\frac{2}{L}\bm{\mu}_{0,2}  +\frac{1}{\lambda}w_{\text{rel}}\cosh(\lambda\tfrac{L}{2})(\bm{\mu}_1^+-\bm{\mu}_1^-)\Big)\nonumber\\
    &=c_{0,1}\bm{\mu}_{\mathcal{B}} -\frac{2}{\lambda}\frac{w_{\mathcal{B}}}{w_{\mathcal{J}}}w_{\text{rel}}\sinh(\lambda\tfrac{L}{2})(\bm{\mu}_1^++\bm{\mu}_1^-).
    \label{eq:mu_j-decomp}
    \end{align}
The equality of both decompositions in \eqref{eq:mu_j-decomp} implies,
\begin{equation}
\bm{\mu}_{\mathcal{B}}\wedge\Big(w_{\text{eq}}\frac{2}{L}\bm{\mu}_{0,2}  +\frac{1}{\lambda}w_{\text{rel}}\big(\cosh(\lambda\tfrac{L}{2})(\bm{\mu}_1^+-\bm{\mu}_1^-) - \sinh(\lambda\tfrac{L}{2})(\bm{\mu}_1^++\bm{\mu}_1^-)\big)\Big)=0
\end{equation}
Replacing by the expression of $\bm{\mu}_{\mathcal{B}}$ \eqref{eq:mub_normed},
$\bm{\mu}_{0,2}$ \eqref{eq:mu02} and $\bm{\mu}^+_1 \pm \bm{\mu}^-_1$  \eqref{eq:mu1pm}, we obtain,
\begin{equation}
  w_{\text{rel}} = \frac{\tilde{w}_{\text{rel}} }{\tilde{w}_{\text{rel}}  + \tilde{w}_{\text{eq}} } \quad \text{and} \quad 
   w_{\text{eq}} = \frac{\tilde{w}_{\text{eq}}}{\tilde{w}_{\text{rel}}  + \tilde{w}_{\text{eq}}} ,
\end{equation}
with 
\begin{align}
  \tilde{w}_{\text{rel}}&= \frac{\omega_2}{\omega_{1,3}}\frac{2\omega_{1,3}+\omega_2}{\omega_{1,3}(1-q_{1,3})+\omega_2}\frac{1-q_{1,3}}{1+q_{1,3}}\nonumber\\
  \tilde{w}_{\text{eq}}                        &=  \frac{\lambda L}{\tanh\big(\lambda\tfrac{L}{2}\big)}\frac{\omega_2+2\omega_{1,3}(1-q_{1,3})}{2\omega_{1,3}+\omega_2} +\lambda L\frac{2\lambda|v_{1,3}-v_2|}{(1+q_{1,3})(2\omega_{1,3}+\omega_2)}.\nonumber
\end{align}
Note that for $q_{1,3} \to 1$ or instant tumble
thermalization/diffusive limit ($L\lambda \gg 1$), we recover the
detailed-jamming scenario ($w_{\text{eq}} \to 1$) and the relaxation
behavior becomes negligible.

By positivity argument, it is required that, at any $r$,
$ \mu(r,+1,2) \propto \tilde{w}_{\text{rel}}\gamma(r,+1,2) +
\tfrac{2}{L}\tilde{w}_{\text{eq}}\mu_{\mathcal{B}}(+1,2) \geq 0$, i.e. $\tilde{w}_{\text{rel}}\leq
\tilde{w}_{\text{eq}} \tfrac{2}{L}\frac{\mu_{\mathcal{B}}(+1,2)}{ |\gamma(r,+1,2)|} $. By
monotonicity of the $\cosh$ and $\sinh$ functions, it is equivalent to,

\[
 2q_{1,3}\Big(\cosh\big(\lambda\tfrac{L}{2}\big)(\omega_2+\omega_{1,3}(1-q_{1,3}))
  +\sinh\big(\lambda\tfrac{L}{2}\big)\lambda|v_{1,3}-v_2|\Big) \geq 0,
\]
which is always satisfied. For $q_{1,3}=0$, we obtain
$ \mu(0^+,+1,2) = 0 = \pi(0^+,+1,2)$ and a maximal deviation from
$\tfrac{2}{L}\mu_{\mathcal{B}}(+1,2)$. Furthermore, in the ballistic
limit $\lambda L \ll 1$, we obtain $\tilde{w}_{\text{rel}}\sim \tilde{w}_{\text{eq}}$ and $\mu(r,+1,2) \sim \mu(r,0^+,2) $
for all interdistance $r$, which also yields a constant
$\mu(r,-1,2) \sim \mu(r,0^+,2) $ by the periodicity detailed condition
$(C_P)$. Thus, in the ballistic limit, the system does not relax
from the conditions imposed by the jamming boundary. In the case of
$q_{1,3}=0$, this interestingly gives $\pi(r,\pm 1,2) \sim 0$ for all
interdistance $r$. \medskip

Finally, the value of the coefficient
$w_{\mathcal{B}}=1-w_{\mathcal{J}}$ is constrained by the
transformation of the bulk probability flows into jamming ones
(cf. the decomposition \eqref{eq:mu_j-decomp}), which should be
normalized, i.e.
$\int_{\mathcal{V}}\d\mu_{\mathcal{J}}(\nu,s)=1$. When expliciting
$\mu_{\mathcal{J}}(0,0)$ and $\mu_{\mathcal{J}}(0,2)$ by their
decomposition into $\mu_{\mathcal{J}}(-1,1)$ and
$\mu_{\mathcal{J}}(-1,2)$ via $(C_{\mathcal{J}})$ (or
\eqref{eq:muj0-decomp}), the normalization conditions writes,
\begin{equation}
\mu_{\mathcal{J}}(-1,1)\Big(1 + \frac{\omega_{1,3}}{2\omega_2}(1-q_{1,3}) + \frac{\omega_2}{4\omega_{1,3}}\Big)+ \mu_{\mathcal{J}}(-1,2)(1+q_{1,3}) \!=\! 1.
\end{equation}
We obtain by replacing $\mu_{\mathcal{J}}(-1,1)$ and
$\mu_{\mathcal{J}}(-1,2)$ by the expression given by the decomposition
\eqref{eq:mu_j-decomp},
\begin{equation}
  w_{\mathcal{B}} = \frac{\tilde{w}_{\mathcal{B}}}{\tilde{w}_{\mathcal{B}}  + \tilde{w}_{\mathcal{J}}} \quad \text{and} \quad 
   w_{\mathcal{J}} = \frac{\tilde{w}_{\mathcal{J}}}{\tilde{w}_{\mathcal{B}}  + \tilde{w}_{\mathcal{J}}} ,
\end{equation}
with 
\begin{align}
  \tilde{w}_{\mathcal{B}}&=\sinh\big(\lambda \tfrac{L}{2}\big)(\tilde{w}_{\text{eq}}+\tilde{w}_{\text{rel}})\frac{\omega_{1,3}\lambda|v_{1,3}-v_2|}{2(\omega_{1,3}(1-q_{1,3})+\omega_2)^2} \\
   \tilde{w}_{\mathcal{J}}&=
  \sinh\big(\lambda\tfrac{L}{2}\big)
\frac{2\omega_{1,3}+\omega_2}{2\lambda|v_{1,3}-v_2|}  
  \Big[
1
  + \frac{\omega_2^2(1+q_{1,3})}{2(\omega_{1,3}(1-q_{1,3})+\omega_2)^2}
  \Big]
 +2\cosh\big(\lambda\tfrac{L}{2}\big).\nonumber
\end{align}
This yields,
\begin{align}
  \frac{w_{\mathcal{J}}}{w_{\mathcal{B}}} =\frac{1}{\tilde{w}_{\text{eq}}+\tilde{w}_{\text{rel}}}&\Big(                                                           
  \frac{2(\omega_{1,3}(1-q_{1,3})+\omega_2)}{\omega_{1,3}(1-q_{1,3})}
  \Big[1  + \frac{\omega_2^2(1+q_{1,3})}{2(\omega_{1,3}(1-q_{1,3})+\omega_2)^2}
  \Big]
 &+\frac{2}{\tanh\big(\lambda\tfrac{L}{2}\big)}\frac{2(\omega_{1,3}(1-q_{1,3})+\omega_2)^2}{\omega_{1,3}\lambda|v_{1,3}-v_2|}\Big).
\end{align}

In the diffusive limit $\lambda L \gg 1$, as
$\tilde{w}_{\text{eq}}+\tilde{w}_{\text{rel}}\gg 1$, we obtain that
$w_{\mathcal{B}} \to 1$, $w_{\mathcal{J}} \to 0$. In such limit, the
stationary distribution $\pi$ comes down to a product measure with no
Dirac masses on the jamming configurations but a uniform distribution
in the bulk. Thus, the marginal of $\pi$ over the dynamical variables corresponds to
the invariant measure of an equilibrium reflected diffusion.  In the
ballistic limit $\lambda L \ll 1$ however, as
$\tilde{w}_{\text{eq}}+\tilde{w}_{\text{rel}}\sim 1$, we obtain that
$w_{\mathcal{B}} \to 0$, $w_{\mathcal{J}} \to 1$. Indeed, in such
limit where the tumble mechanism becomes negligible, the system cannot
escape the jamming configuration and yields, once marginalized over
the dynamical variables, the same invariant measure as some RTP without refreshment. This is
 consistent with the interpretation of the quantity $\lambda L$
 as a measure of ballisticity vs diffusivity.

\subsection{Asymmetrical three-state RTP systems and  two-eigenvalue relaxation}
\label{sec:app-anisotropic}

\begin{figure}
     \centering
     \raisebox{1.5\height}{\includegraphics[width=0.3\textwidth]{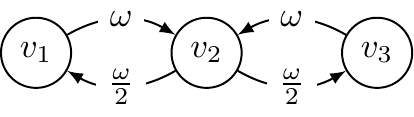}}
     \hfill
         \includegraphics[width=0.6\textwidth]{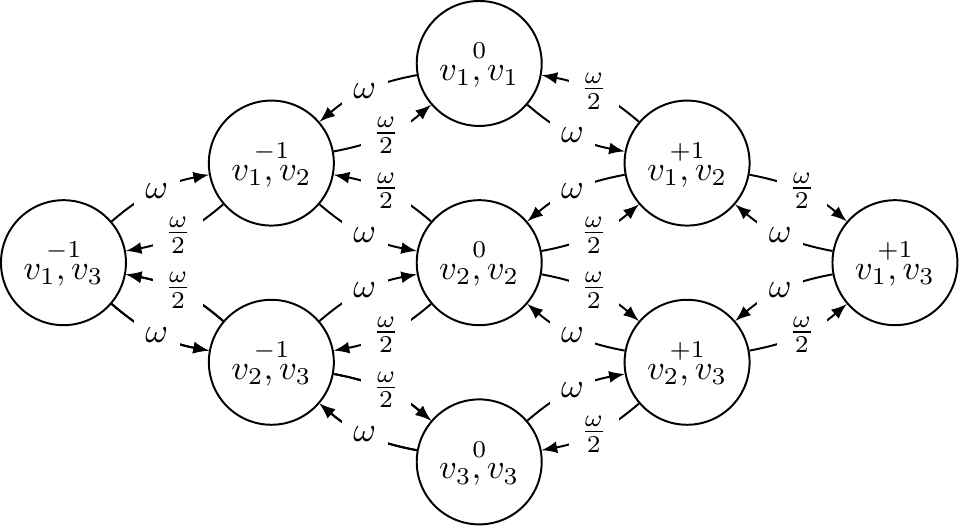} 
\caption{Transition rates (i.e. $\omega(\cdot)q(\cdot,\cdot)$)
   between the internal states of a single anisotropic 3-state RTP
   ({\bf left}) and corresponding transition rates
   (i.e. $\tilde{\omega}(\cdot)Q(\cdot,\cdot)$) at the interdistance
   level in the representation
   $s_t = \{\{v_{t,1}, v_{t,2}\}\}$~({\bf right}).}
   \label{fig:anisotropic_rate}
\end{figure}

We now illustrate how breaking the symmetry of the previous
three-state RTP model effectively leads to the appearance of multiple
lengthscales and exponentials in the bulk, as written in
\eqref{eq:main_cat_sol}. To do so, we introduce some asymmetry in the
velocity reached by a RTP in its different states
(e.g. $|v_1-v_2|\neq|v_3-v_2|$). As illustrated in
Figure~\ref{fig:anisotropic_rate}, all other parameters are kept
symmetrical. This situation can be for instance realized in the case
of an anisotropy in the velocity reached by a RTP given its direction
(i.e. $v_1 = -2$, $v_2 = 0$ and $v_3 = 1$). At the interdistance level, we use
the representation through multisets $s_t = \{\{v_{t,1},
v_{t,2}\}\}$. It corresponds to the set of internal states
$\mathcal V = \mathcal V_{-1} \cup \mathcal V_0 \cup \mathcal V_{1}$
where
$$
\mathcal V_{\pm 1} = \{\pm 1\} \times \{\{v_2,v_1\},\{v_2,v_3\},\{v_1,v_3\}\} \text{ and } \mathcal V_0 = \{0\} \times \{\{v_1,v_1\}, \{v_2,v_2\}, \{v_3,v_3\}\}
$$
with the transition rates of Figure~\ref{fig:anisotropic_rate}. The tumble invariant measure is then,
\begin{align*}
\mu_{\mathcal B}(\nu, s) &= 2^{m_s(v_2)}/16,
\end{align*}
with $m_s(\cdot)$ the multiplicity of an element in $s$.

Without loss of generality, we consider $|v_1-v_2|>|v_2-v_3|$ and note $${\rho} = (v_3-v_2)/(v_1-v_2),$$ the asymmetry factor ($0<|{\rho}|<1$). Similarly to Section \ref{sec:app-3rtp}, we obtain,
       $$M^+ = \bordermatrix{ &  s = \{v_1,v_2\} & s = \{v_2,v_3\} & s=\{v_3,v_1\}   \cr
	s = \{v_1,v_2\} & -2|{\rho}|(1-{\rho}) & |{\rho}|(1-{\rho}) & 2|{\rho}|(1-{\rho}) \cr
		      &      &            &                       \cr
	s = \{v_2,v_3\} &  (1-{\rho}) &-2(1-{\rho}) & 2(1-{\rho}) \cr 
		      &     &             &                      \cr
	s = \{v_3,v_1\} & |{\rho}| &  |{\rho}| &  -4|{\rho}| \cr } \frac{\omega}{2|{\rho}||v_1-v_3|}
      $$
 and,
 $$M^- = \bordermatrix{ &  s = \{v_1,v_2\} & s = \{v_2,v_3\} & s=\{v_3,v_1\}   \cr
	s = \{v_1,v_2\} & -4|{\rho}|(1-{\rho})  & 0 &2|{\rho}|(1-{\rho}) \cr
		      &      &            &                       \cr
	s = \{v_2,v_3\} & 0 & -4(1-{\rho}) & 2(1-{\rho})\cr 
		      &     &             &                       \cr
	s = \{v_3,v_1\} &  |{\rho}|&|{\rho}| &-4|{\rho}| \cr }\frac{\omega}{2|{\rho}||v_1-v_3|}.
      $$
      The matrices $M^\pm$ are not collapsable anymore, due to the
      asymmetry. We now study the impact of the asymmetry on the
      spectrum properties of $M^\pm$ and on the bulk relaxation
      behavior. Given the dimensionality and the fact that $0$ is an
      eigenvalue, $M^+M^-$ admits two other eigenvalues
      $\lambda^2_\pm$ we now determine. We obtain,
  
      $$\text{Tr}(M^+M^-)=\frac{\omega^2}{{\rho}^2|v_1-v_3|^2}T({\rho}),
      $$
      with
      $$T({\rho}) = 2(1+{\rho}^2)(1-{\rho})^2 + |{\rho}|(1+|{\rho}|)(1-{\rho}) + 4{\rho}^2 >0$$
and,     
      \begin{multline*}
        \text{det}(M^+M^--\lambda^2\text{Id}) =
        -\lambda^2\Big(\lambda^4-\lambda^2\text{Tr}(M^+M^-)
        +\frac{\omega^4(1-{\rho})^2}{4|v_1-v_3|^4{\rho}^2} \Big[
          12(1-{\rho})(2-{\rho}+|{\rho}|) 
        + 21(1+{\rho}^2)
        + 6|{\rho}|\Big]\Big).
      \end{multline*}
      Therefore, the eigenvalues are,
      \begin{equation}
        \lambda^2_\pm = \frac{\omega^2}{{\rho}^2|v_1-v_3|^2} \Big(\frac{T({\rho})}{2}\pm (1+{\rho})\sqrt{\Delta(r)}\Big),
      \end{equation}
      with,
      \begin{equation*}
        \Delta({\rho})=\left\{
      \begin{array}{ll}
        \Big({\rho}^6- 7{\rho}^5 +18{\rho}^4 -23{\rho}^3+18{\rho}^2-7{\rho}+1\Big)>0 &\text{if ${\rho}<0$}\\
      \frac{1}{(1+{\rho})^2}\Big({\rho}^8-5{\rho}^7+9{\rho}^6-8{\rho}^5  + 10{\rho}^3 - {\rho}^2 -3{\rho}+1\Big)>0 &\text{if ${\rho}>0$}
        \end{array}\right.,
    \end{equation*}
so that, $0 \leq \sqrt{\Delta} \leq 2$.
    
Thus, such asymmetrical three-state RTP systems always exhibit a
relaxation in the bulk over two distinct lengthscales. The limit value
${\rho}\to-1$, belonging to the previous symmetrical subclass, recovers the
single lengthscale relaxation. Interestingly, the other limits ${\rho}=0,1$
lead to the divergence of the relaxation lengthscales. This is
consistent with the fact that such limits actually correspond to a
detailed-jamming two-state RTP system. 
Finally, for $0<|{\rho}|<1$, the relaxation terms are
\begin{equation}
  \begin{aligned}
  \Big( \begin{array}{l}
          \gamma(r,\pm 1, \{v_1,v_2\}) \\
          \gamma(r,\pm 1,\{v_2,v_3\})
        \end{array}\Big)
        \!=\!                            
            \frac{1}{\mathcal{N}_\gamma}\big[&b_1\cosh\Big(\lambda_+\Big(\frac{L}{2}\!-\!r\Big)\Big) 
          (\bm{\mu}_{1}^+  \!+\! \bm{\mu}_{1}^-) 
    \!\mp b_1\sinh\Big(\lambda_+\Big(\frac{L}{2}\!-\!r\Big)\Big)
    ( \bm{\mu}_{1}^+   \!-\!\bm{\mu}_{1}^-)\\
    &+b_{-1}\cosh\Big(\lambda_-\Big(\frac{L}{2}\!-\!r\Big)\Big) 
          (\bm{\mu}_{-1}^+  \!+\! \bm{\mu}_{-1}^-) 
    \!\mp b_{-1}\sinh\Big(\lambda_-\Big(\frac{L}{2}\!-\!r\Big)\Big)
    ( \bm{\mu}_{-1}^+   \!-\!\bm{\mu}_{-1}^-)\big].
    \end{aligned}
\label{eq:asym-3-state-gamma}  
\end{equation}

\begin{figure}[!h]
	\centering
	\includegraphics[width=.7\textwidth]{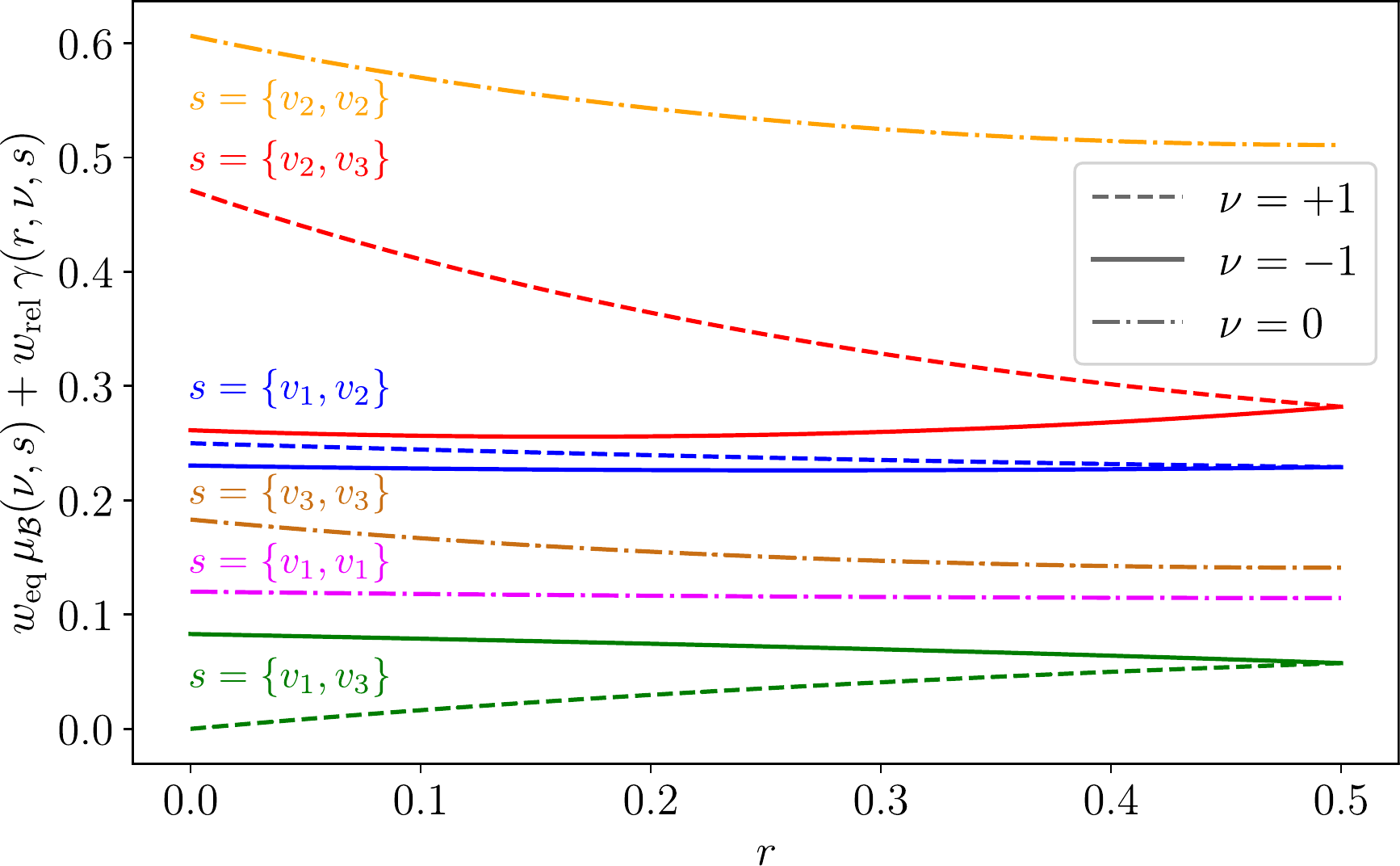}
	\caption{Density of the invariant measure in the bulk for $\omega=1.5$ and $L=1$}
	\label{fig:anisotropic_densities}
\end{figure}

\begin{figure}[!h]
	\centering
	\includegraphics[width=.7\textwidth]{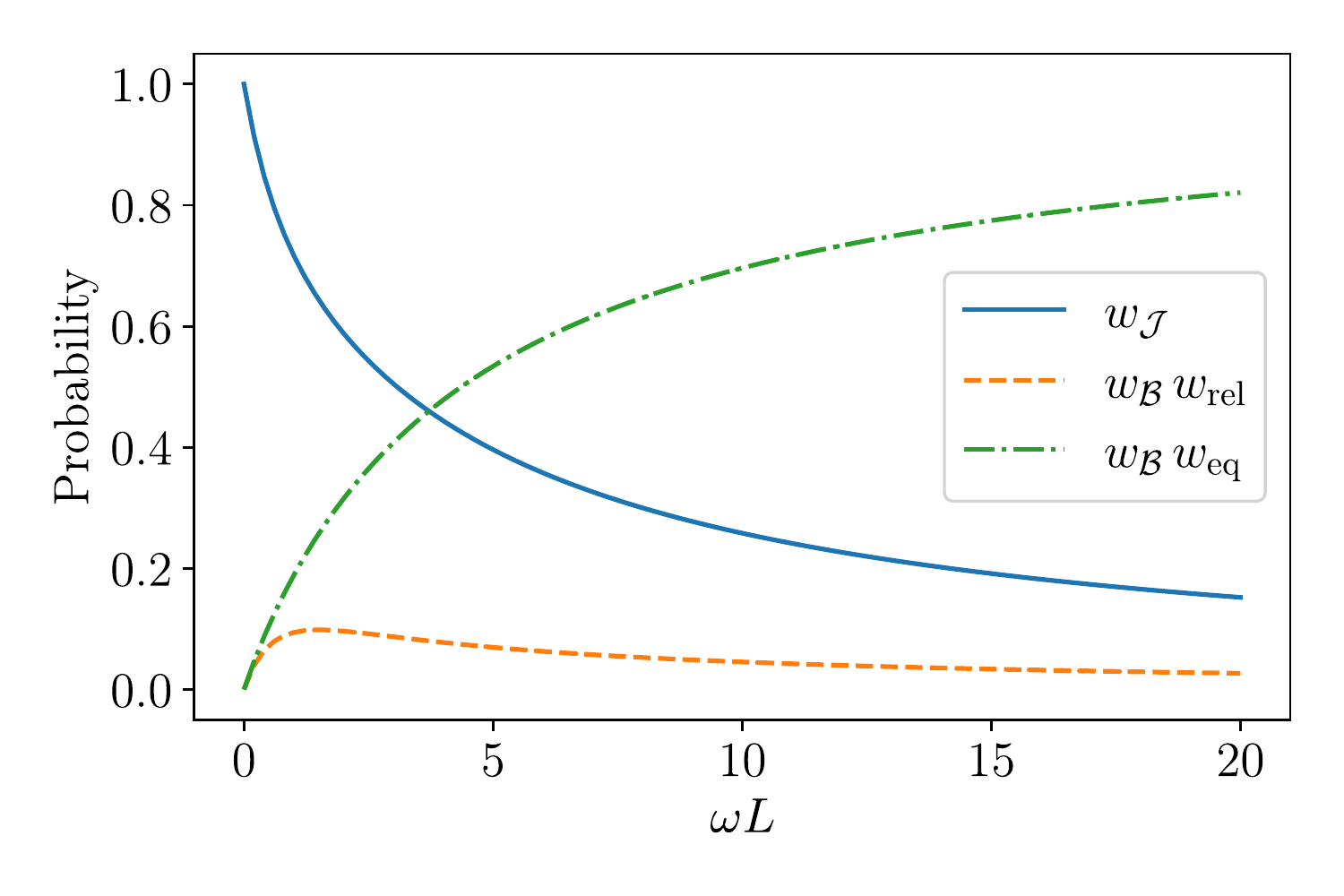}
	\caption{Respective contributions of the Dirac, exponential and uniform terms towards the invariant measure as a function of $\omega L$}
	\label{fig:dirac_expo_unif_mass_plot}
\end{figure}

Similarly to the symmetrical three-state RTP case, the nature of the
regime, either diffusive or ballistic is ruled by $\lambda_{+}L$ or
$\lambda_- L$, which both scale as $\omega L/({\rho}|v_1-v_3|)$.  We
study the impact of such regimes in the particular case
$v_1=-2, v_2=0, v_3=1$. We obtain
$$
\lambda_+ = \frac{\omega}{6}  \sqrt{62 + 11  \sqrt{7}}, \quad \lambda_- = \frac{\omega}{6}  \sqrt{62 -11  \sqrt{7}}
$$
and  the
  following explicit expression for $\gamma$,
\begin{align*}
  \begin{pmatrix}
 \gamma(r,0,\{ v_3, v_3 \}) \\
 \gamma(r,0,\{ v_2, v_2\}) \\
 \gamma(r,0,\{ v_1, v_1 \})
\end{pmatrix}
&= \frac{1}{N_\gamma} \sum_{\sigma\in\{1,-1\}}  b_\sigma
\cosh\left( \lambda_\sigma \left(r - L/2\right) \right)
\begin{pmatrix}
227 + 86 \sigma \sqrt{7} \\
3(148 + 55 \sigma \sqrt{7}) \\
- \frac{1}{2}(10 + 7 \sigma \sqrt{7})
\end{pmatrix}, \\
  \begin{pmatrix}
 \gamma(r,\pm 1, \{v_2, v_3\}) \\
 \gamma(r,\pm 1,\{ v_1, v_3 \}) \\
 \gamma (r, \pm 1,\{ v_2, v_3\} ) \\
\end{pmatrix}
&= \frac{1}{N_\gamma}
\sum_{\sigma\in\{1,-1\}} b_\sigma\Bigg [
\cosh\left(  \lambda_\sigma \left(r - L/2\right) \right)
\begin{pmatrix}
2(227+86\sigma\sqrt{7}) \\
-\frac{1}{6}(355+127\sigma\sqrt{7}) \\
-(10+7\sigma\sqrt{7})\\
\end{pmatrix} \\
&\quad  \mp  \lambda_\sigma\sinh\left( \lambda_\sigma \left(r - L/2\right) \right)
\begin{pmatrix}
6(34 + 13\sigma \sqrt{7}) \\
-2(23 + 8\sigma\sqrt{7}) \\
-3(11 + 5\sigma\sqrt{7}) \\
\end{pmatrix}\Bigg].
\end{align*}
We determined 
a complete
characterization of the invariant measure $\pi$, as displayed in
Figure~\ref{fig:anisotropic_densities}. This enables a comparison of
the weight of the jamming Dirac terms $w_{\mathcal J}$, the relaxation
terms $w_{\mathcal B} \, w_\text{rel}$ and the equilibrium terms
$w_{\mathcal B} \, w_\text{eq}$.  The evolution of these quantities
depending on the regime being diffusive or ballistic, as set by
$\omega L$, can be found in Figure~\ref{fig:dirac_expo_unif_mass_plot}
and is similar to the evolution observed for the symmetrical
three-state RTP systems previously discussed. Furthermore, we now give
the Taylor expansions of $w_\mathcal J$,
$w_{\mathcal B} \, w_\text{rel}$ and $w_{\mathcal B} \, w_\text{eq}$
when $\omega L \ll 1$
\begin{align*}
w_\mathcal J  &= 1 -\frac{63}{128} {\omega L} + O((\omega L)^2), \\
w_{\mathcal B} \, w_\text{rel} &= \frac{1179}{4736} {\omega L} + O((\omega L)^2), \\
w_{\mathcal B} \, w_\text{eq} &= \frac{9}{37} {\omega L} + O((\omega L)^2),
\end{align*}
and when $\omega L \gg 1$
\begin{align*}
w_\mathcal J &= \frac{1}{96} \left(148+\sqrt{\frac{37}{7} \left(4675+579 \sqrt{37}\right)}\right) \frac {1}{\omega L} + O\left(\frac{1}{(\omega L)^2}\right),\\
w_{\mathcal B} \, w_\text{rel} &= \left(\frac{107}{24}-\frac{1}{96} \sqrt{\frac{17631787}{259}+\frac{75087 \sqrt{37}}{7}}\right) \frac {1}{\omega L} + O\left(\frac{1}{(\omega L)^2}\right), \\
w_{\mathcal B} \, w_\text{eq} &= 1 + \left(\sqrt{\frac{355}{259}+\frac{3 \sqrt{37}}{14}}-6\right) \frac{1}{\omega L} + O\left(\frac{1}{(\omega L)^2}\right).
\end{align*}
Similarly to the symmetrical 3-RTP systems
discussed previously, these expansions show that in the ballistic
limit $\lambda_{\pm}L \rightarrow 0$ the invariant measure collapses
to Dirac masses on the jamming configurations, whereas in the
diffusive $\lambda_\pm L \rightarrow +\infty$ limit the invariant
measure converges towards a bulk uniform distribution with no Dirac
masses.

\section{Discussion}
\label{sec:discuss}

The approach presented here gives a direct continuous-time and
continuous-space setting and encodes the fundamental symmetries
(homogeneity, particle indistinguishability, periodicity). By
adequately capturing the jamming behavior as some boundary effect on
the bulk, we are able to define the two universality classes of the
steady state. Interestingly, they are entirely defined from the
satisfaction or not of a detailed dynamical symmetry (\ref{eq:DB}) at
jamming. Such symmetry translates in the bulk into the satisfaction of
the necessary active global balance (\ref{eq:all_flow}) in a stricter
--and closer to equilibrium-- detailed manner
(\ref{eq:det_flow}). Then, the further away from the detailed symmetry
(\ref{eq:DB}) the jamming scenario constrains the system, the more
involved and rich relaxation behavior are
eventually exhibited in the bulk, following (\ref{eq:edo}). Therefore,
the system is analogous to a catenary problem. The active global
balance acts as the fixed-length constraint on a cable --or equivalently
the RTP activity acts as some tension-- and the tumble process acts as a
weight force.

Thus, apart from the Dirac terms, the impact of the time-reversibility
breaking by the activity simply translates itself into the capacity of
the system to explore its boundaries. There, the now \emph{seen}
jamming interactions may further impact the steady state, depending on
some symmetry preservation details. It is interestingly reminiscent of
the situation in non-reversible sampling by Event-Chain Monte Carlo
\cite{bernard09, Michel_2014}, which generates in particle systems
continuous-time Markov processes coming down to an artificial RTP
dynamics, also encapsulated as PDMP \cite{monemvassitis23}. Time
reversibility is broken but the equilibrium Boltzmann distribution is
still left invariant by the preservation at collisions of some key
symmetries, be it pairwise~\cite{Michel_2014},
translational~\cite{Harland_2017} or rotational~\cite{Michel_2020}.
For systems in the global-jamming class, breaking the detailed
dynamical symmetry at jamming, we show for different cases that the
relaxation lengthscale $\lambda$ sets the dimensionless quantity
$\lambda L$, separating diffusive ($\lambda L \gg 1$) from ballistic
($\lambda L\ll 1$) regimes. In the diffusive regime, an equilibrium
behavior is recovered, while, in the ballistic one, the system cannot
escape jamming configurations.

As discussed, the presented conclusions generalize
to more general jamming scenario $\mathcal{A}_J$, set by some
tumbling rates and kernel different from the bulk, as it only impacts
the exact form of the weights  and
$\mu_{\mathcal{J}}(\cdot)$. Some future work lies in the study of the
impact of non-homogeneity, with the tumbling depending on the RTP
positions, as in presence of an external potential or soft
interactions. The question of the relaxation to stationarity should
also find in the PDMP formalism a powerful approach that should be
explored. More importantly, a compelling research perspective lies in
the use of PDMPs and their faculty to isolate the impacts of boundary
effects on the bulk to model more than two RTPs with hardcore jamming
interactions. An explicit formula or a good quantitative understanding
of the invariant measure would provide a new perspective on clustering
phenomena such as MIPS, with a direct understanding of the impact of
the microscopic details.\\

\begin{acknowledgments}
  All the authors acknowledge the support of the French ANR under the
  grant ANR-20-CE46-0007 (\emph{SuSa} project).  This work has also been
  (partially) supported by the French ANR under the grant ANR-23-CE40-0003
  (\emph{CONVIVIALITY} Project).
\end{acknowledgments}

\bibliography{biblio}

\appendix

\section{Global-jamming steady state: Spectral analysis of bulk matrix $B$}
\label{app:spectrum}

We study in this Section the
spectral properties of $B$, which are strongly constrained by
the symmetry (\ref{eq:sym_Q}) of $Q$. 

\subsection{Jordan reduction of $B$ and symmetry constraints}

We now consider the
Jordan reduction of $B = P^{-1}JP$, with $J$ a Jordan matrix. For
the Jordan block $k$ of size $p_k$, we note the corresponding
eigenvalue $\lambda_k$ and the Jordan chain of linearly-independent
generalized eigenvectors
$\bm{\mu_{k,n}} =(\bm{\mu}^+_{k,n},\bm{\mu}^-_{k,n})$, with
$1\leq n \leq p_k$, $\bm{\mu_{k,1}}$ being an eigenvector and
$B\bm{\mu}_{k,n} = \lambda_k\bm{\mu}_{k,n} + \bm{\mu}_{k,n-1}$ for $n>1$.

By the symmetry of $B$, any block $k$ of
eigenvalue $\lambda_k > 0$ is mirrored by a block $k^*$ of
eigenvalue $\lambda_{k^*}=-\lambda_k$ and generalized eigenvectors
$\bm{\mu}_{k^*,n} =
(-1)^{n+1}  (\bm{\mu}^-_{k,n},\bm{\mu}^+_{k,n})$, as  $B(\bm{\mu}^-_{k,1},\bm{\mu}^+_{k,1})=-\lambda_k(\bm{\mu}^-_{k,1},\bm{\mu}^+_{k,1})$  and, for $n>1$,
\[(-1)^{n+1}B(\bm{\mu}^-_{k,n},\bm{\mu}^+_{k,n})= -\lambda_k(-1)^{n+1}(\bm{\mu}^-_{k,n},\bm{\mu}^+_{k,n}) 
+(-1)^{n}(\bm{\mu}^-_{k,n-1},\bm{\mu}^+_{k,n-1}).
\]

Furthermore, as $\hat{D}^\pm_{\phi}B\hat{D}^{-1}_\omega$, with
$\hat{D}^\pm_{\phi}=\left(\begin{smallmatrix}D_{\phi} & 0\\ 0
&-D_{\phi}\\\end{smallmatrix}\right)$ and $\hat{D}^{-1}_{\omega}=\left(\begin{smallmatrix}D^{-1}_{\omega} & 0\\ 0
&D^{-1}_{\omega}\\\end{smallmatrix}\right)$ , identifies with the
tumble Markov kernel collapsed in $\nu=0$ of the considered process, it is
diagonalizable with real eigenvalues ranging from $0$ to $-1$ and a
unique eigenvector $\bm{\mu}$ so that $\hat{D}^\pm_{\phi}B\hat{D}^{-1}_\omega\bm{\mu} =0$. The vector $\bm{\mu}$ identifies with the unique stationary bulk solution
as  $\bm{\mu} = \hat{D}_\omega(\bm{\mu}_{\mathcal{B}},\bm{\mu}_{\mathcal{B}})$. As $\hat{D}_\phi$ is invertible,
$(\bm{\mu}_{\mathcal{B}},\bm{\mu}_{\mathcal{B}})$ is the unique eigenvector of eigenvalue
$0$ of $B$.
Then, for the unique block $k$ with $\lambda_k = 0$, the generalized
eigenvectors are of the following form
$\bm{\mu_{k,n}} =(\bm{\mu}^+_{k,n},(-1)^{n+1}\bm{\mu}^+_{k,n})$, as,
\[\left\{\begin{array}{l} \bm{\mu}_{k,1}=(\bm{\mu}_{\mathcal{B}},\bm{\mu}_{\mathcal{B}})\\ B\bm{\mu}_{k,n} = \bm{\mu}_{k,n-1}, \ n>1 \end{array}\right. \ \ \text{gives} \ \ \left\{\begin{array}{ll}
(B^+ + (-1)^{n+1}B^-) \bm{\mu}^+_{k,n} &=  \bm{\mu}^+_{k,n-1}\\
-((-1)^{n+1}B^+ + B^-) \bm{\mu}^+_{k,n} &= (-1)^{n}\bm{\mu}^+_{k,n-1}\\
\end{array}\right..
\]

It is finally noteworthy that
$(\bm{\mu}_{\mathcal{B}},\bm{\mu}_{\mathcal{B}})$ is the only
symmetrical eigenvector of $B$, as any symmetrical eigenvector
is of eigenvalue $0$, as can be seen directly from $B$ or from
an equivalent argument of linear independence for
$\lambda_k\neq 0$ of the eigenvectors
$(\bm{\mu}^+_{k,n},\bm{\mu}^-_{k,n})$ (e.v. $\lambda_k$) and
$(\bm{\mu}^-_{k,n},\bm{\mu}^+_{k,n})$ (e.v. $-\lambda_k$).
This is consistent with the fact that a symmetrical
eigenvector is by definition a stationary solution of the
detailed-jamming class which is unique and of eigenvalue
$0$. Hence the complete characterization only by the
detailed-symmetry condition ($\pi(r,\nu,s) = -\pi(r,-\nu,s)$)
of the detailed-jamming class, which is equivalent to
$\pi(r,\nu,s) = \mu_Q(\nu,s)$.\\

Thus, the number of Jordan blocks of $J$ is odd and is noted
$2d+1$. We rank the blocks from $-d$ to $d$ in a symmetrical
manner: The $k$-th block of eigenvalue $\lambda_k$ and
generalized eigenvectors $(\bm{\mu}^+_{k,n},\bm{\mu}^-_{k,n})$
is mirrored by the $(-k)$-th block of eigenvalue $-\lambda_k$,
generalized eigenvectors
$((-1)^{n+1}\bm{\mu}^-_{k,n},(-1)^{n+1}\bm{\mu}^+_{k,n})$ .
We then set the position of the block linked to the
eigenvector
$(\bm{\mu}_{0,1},\bm{\mu}_{0,1})=(\bm{\mu}_{\mathcal{B}},\bm{\mu}_{\mathcal{B}})$ to
the $0$-th rank. It is composed of generalized eigenvectors of
the form $(\bm{\mu}_{0,n},(-1)^{n+1}\bm{\mu}_{0,n})$.

It is noteworthy that, as $B$ is of even size, the Jordan
block associated to $\lambda_0 = 0$ must be of an even size
$p_0>0$, so that $B$ is never diagonalizable. This is a direct
consequence of the symmetry imposed on $Q$ by the particle
indistinguishability and space homogeneity.

\subsection{Spectral relationship with submatrices $B_+$ and $B_-$ and catenary equation}        \label{app:submat}
The family
$((\bm{\mu}^+_{k,n},\bm{\mu}^-_{k,n})_{k,n},(\bm{\mu}_{0,n},(-1)^{n+1}\bm{\mu}_{0,n})_n,(\bm{\mu}^-_{k,n},\bm{\mu}^+_{k,n})_{k,n})$
naturally forms a basis of
$\mathbb{R}^{2|\mathcal{V}_1|}$. Furthermore, the vectors
$(\bm{\mu}^+_{k,n}+\bm{\mu}^-_{k,n})_{k,n}$
(resp. $(\bm{\mu}^+_{k,n}-\bm{\mu}^-_{k,n})_{k,n}$) are
linearly independent. Indeed, if it exists $\{b_{k,n}\}$ so
that
$\sum_{k,n} b_{k,n}\bm{\mu}^+_{k,n} = - \sum_{k,n}
b_{k,n}\bm{\mu}^-_{k,n}$ (resp.
$= \sum_{k,n} b_{k,n}\bm{\mu}^-_{k,n}$), then we have
$\sum_{k,n} b_{k,n}\bm{\mu}_{k,n} = -
\sum_{k,n}b_{k,n}(-1)^{n+1}\bm{\mu}_{-k,n}$ (resp.
$= \sum_{k,n} b_{k,n}\bm{\mu}_{k,n}$) , which is not
possible. Eventually, by considering the size of the sets
$(\bm{\mu}^+_{k,n}\pm\bm{\mu}^-_{k,n})$, they form two
basis of $\mathbb{R}^{|\mathcal{V}_1|}$.

In
addition, we have the relationships, noting $M^+ = (B^++B^-)$ and $M^-=(B^+-B^-)$,
\begin{equation}
\begin{aligned}
\text{For $k>0$, }&\left\{\begin{array}{l}
M^+(\bm{\mu}^+_{k,1}+\bm{\mu}^-_{k,1}) = \lambda_k(\bm{\mu}^+_{k,1}-\bm{\mu}^-_{k,1})\\
M^-(\bm{\mu}^+_{k,1}-\bm{\mu}^-_{k,1}) = \lambda_k(\bm{\mu}^+_{k,1}+\bm{\mu}^-_{k,1})
\end{array}\right.&\text{and}
&\left\{\begin{array}{l}
M^+\bm{\mu}_{0,1} = 0\\
M^-\bm{\mu}_{0,2} = \bm{\mu}_{0,1}\\
\end{array}\right.                              
\end{aligned},
\label{eq:vec}
\end{equation}
so that, the vectors $(\bm{\mu}^+_{k,1}\pm\bm{\mu}^-_{k,1})_{k>1}$
are respectively the eigenvectors of $M^\mp M^\pm$ of eigenvalues
$\lambda_k^2$ and $\bm{\mu}_{0,1}$ (resp.  $\bm{\mu}_{0,2}$) is the
eigenvector of $M^-M^+$ (resp. $M^+M^-$) of eigenvalue $0$.

And furthermore, for $n>1$, we have, 
\begin{equation}
\begin{aligned}       
\text{For $k>0$, }&\left\{\begin{array}{l}
M^+(\bm{\mu}^+_{k,n}+\bm{\mu}^-_{k,n}) = \lambda_k(\bm{\mu}^+_{k,n}-\bm{\mu}^-_{k,n})+(\bm{\mu}^+_{k,n-1}-\bm{\mu}^-_{k,n-1})\\
M^-(\bm{\mu}^+_{k,n}-\bm{\mu}^-_{k,n}) = \lambda_k(\bm{\mu}^+_{k,n}+\bm{\mu}^-_{k,n})+(\bm{\mu}^+_{k,n-1}+\bm{\mu}^-_{k,n-1})\end{array}\right.
\\\text{and}
&\left\{\begin{array}{l}
M^+\bm{\mu}_{0,2n+1} = \bm{\mu}_{0,2n}\\
M^-\bm{\mu}_{0,2n} = \bm{\mu}_{0,2n-1}.
\end{array}\right.
\end{aligned}
\label{eq:genvec1}
\end{equation}
Giving, for $p<n$,
\begin{equation}
\begin{aligned}
\text{For $k>0$, }&  \left\{\begin{array}{l}
(M^-M^+-\lambda_k^2)^{p}(\bm{\mu}^+_{k,n}+\bm{\mu}^-_{k,n}) \neq \bm{0}\\
(M^+M^--\lambda_k^2)^{p}(\bm{\mu}^+_{k,n}-\bm{\mu}^-_{k,n}) \neq \bm{0}\end{array}\right. \\
\text{ and }&  \left\{\begin{array}{l}
(M^-M^+)^{p}\bm{\mu}_{0,2n+1} \neq \bm{0}\\
(M^+M^-)^{p}\bm{\mu}_{0,2n}\neq \bm{0} \end{array}\right.  \\
\text{For $k>0$, }&  \left\{\begin{array}{l}
(M^-M^+-\lambda_k^2)^{n-1}(\bm{\mu}^+_{k,n}+\bm{\mu}^-_{k,n}) = \bm{\mu}^+_{k,1}+\bm{\mu}^-_{k,1}\\
(M^+M^--\lambda_k^2)^{n-1}(\bm{\mu}^+_{k,n}-\bm{\mu}^-_{k,n}) = \bm{\mu}^+_{k,1}-\bm{\mu}^-_{k,1} \end{array}\right.             \\
\text{ and } &
\left\{\begin{array}{l}
(M^-M^+)^{n-1}\bm{\mu}_{0,2n-1} =\bm{\mu}_{0,1}\\
(M^+M^-)^{n-1}\bm{\mu}_{0,2n}= \bm{\mu}_{0,2},\end{array}\right.                             
\end{aligned}
\label{eq:genvec2}
\end{equation}
so that,
$((\bm{\mu}^+_{k,n}+\bm{\mu}^-_{k,n})_{k,1<n\leq p_k},
(\bm{\mu}_{0,2n-1})_{1<n\leq p_0/2})$ (resp.
$((\bm{\mu}^+_{k,n}-\bm{\mu}^-_{k,n})_{k,1<n\leq p_k},
(\bm{\mu}_{0,2n})_{1< n\leq p_0/2})$) are generalized eigenvectors
of $M^-M^+$ (resp. of $M^+M^-$).

More generally, the spectral properties of $B$ and $M^+M^-$, $M^-M^+$
are strongly related. Indeed, the characteristic polynomial of $B$
writes itself in terms of $B^2$, as,
\begin{equation*}
\begin{aligned}
B^{p_0}\prod_{k=1}^d(B-\lambda_kI)^{p_k}(B+\lambda_kI)^{p_k}= (B^2)^{p_0/2}\prod_{k=1}^d(B^2-\lambda_k^2I)^{p_k} &=0\\    
B^{2(p_0/2)} \sum_{m=0}^{d}B^{2m}\sum_{\{i_k\}_{k=1}^{d-m}}\prod_{k=1}^{d-m}\lambda_{i_k}^2&=0.   \\
\end{aligned}
\end{equation*}
And, as,
\[  B^{2m} 
= \frac{1}{2}\left( \begin{array}{cc}
(M^-M^+)^{m} + (M^+M^-)^{m}  &   (M^-M^+)^{m} - (M^+M^-)^{m}\\
(M^-M^+)^{m} - (M^+M^-)^{m} &     (M^-M^+)^{m} + (M^+M^-)^{m}
\end{array}\right),
\]
it is equivalent to,
\begin{equation*}
\left\{   \begin{array}{rl}
(M^+M^-)^{p_0/2}\sum_{m=0}^{d}(M^+M^-)^{m}\sum_{\{i_k\}_{k=1}^{d-m}}\prod_{k=1}^{d-m}\lambda_{i_k}^2&=0   \\
(M^-M^+)^{p_0/2}\sum_{m=0}^{d} (M^-M^+)^{m}\sum_{\{i_k\}_{k=1}^{d-m}}\prod_{k=1}^{d-m}\lambda_{i_k}^2&=0.   \\
\end{array}\right.
\end{equation*}
So that, the characteristic polynomials of $M^+M^-$ and $M^-M^+$ are,
\begin{equation*}
(M^\pm M^\mp)^{p_0/2}\prod_{k=1}^d(M^\pm M^\mp-\lambda_k^2I)^{p_k}=0.\\    
\end{equation*}
A similar derivation can be carried out regarding the minimal
polynomial and, thus, the fact that $B$ admits no generalized
eigenvectors apart the necessary $\bm{\mu}_{0,2}$ (i.e. $p_0=2$,
$p_{k\neq 0} = 1$) is equivalent to $M^+ M^-$ and $M^- M^+$
both diagonalizable.

\section{Global-jamming steady state: General solution}
\label{app:gen_bulk}

\subsection{Bulk general form}
\label{app:bulk}
Exploiting the Jordan form of $B$, the solution to Eq.~\ref{eq:edo}
is,
\begin{equation}
\begin{array}{rl}
\left(\begin{array}{c}
\bm{\gamma}^+\\
\bm{\gamma}^-\\
\end{array}\right)(r) &=
\sum_{n=1}^{p_0}\sum_{l=n}^{p_0}a_{0,l} \frac{r^{l-n}}{(l-n)!}
\left(\begin{array}{c}
\bm{\mu}_{0,n}   \\
(-1)^{n+1}\bm{\mu}_{0,n} \\
\end{array}\right)\\
&  \qquad  + \sum_{k=1}^{d} \Big[ \exp\left(-\lambda_k\left(\frac{L}{2}-r\right)\right)\sum_{n=1}^{p_k}\sum_{l=n}^{p_k}a_{k,l} \frac{r^{l-n}}{(l-n)!}
\left(\begin{array}{c}
\bm{\mu}_{k,n}^+   \\
\bm{\mu}_{k,n}^- \\
\end{array}\right)\\
&\qquad + \exp\left(\lambda_k\left(\frac{L}{2}-r\right)\right) \sum_{n=1}^{p_k}\sum_{l=n}^{p_k}a_{-k,l} \frac{r^{l-n}}{(l-n)!}(-1)^{n+1}
\left(\begin{array}{c}
\bm{\mu}_{k,n}^-   \\
\bm{\mu}_{k,n}^+ \\
\end{array}\right)\Big],     
\end{array}
\label{eq:sol}  
\end{equation}
with $(a_{\pm k,l})_{k,l}$ real constants and $a_{0,1}=0$ (no uniform component along 
$(\mu_{\mathcal{B}},\mu_{\mathcal{B}})$). It is
straightforward to check that if $\bm{\gamma}^+(r) = \bm{\gamma}^-(r)$
for all $r$, then $\bm{\gamma}^+(r) = \bm{\gamma}^-(r)=0$ and we
recover the detailed-jamming case. We can also check for the active
global balance,
\begin{equation}
\begin{array}{l}
\sum_s\phi(s)(\bm{\gamma}^+(r)- \bm{\gamma}^-(r))_s = 0,\\
\displaystyle\sum_{k=1}^{d}\sum_{1\le n\le l\le p_k}\frac{r^{l-n}}{(l-n)!}
\big(a_{k,l}e^{-\lambda_k(\frac{L}{2} - r)} - (-1)^{n+1}a_{-k,l} e^{\lambda_k(\frac{L}{2} - r)}\big)  \\
\displaystyle \qquad\times \sum_s\phi(s)(\bm{\mu}_{k,n}^+(s)  - \bm{\mu}_{k,n}^-(s)) +  \sum_{n=1}^{p_0/2}\sum_{l=2n}^{p_0}\frac{r^{l-2n}}{(l-2n)!}2a_{0,l} \sum_s\phi(s)\bm{\mu}_{0,2n}(s)=0.
\end{array}
\label{eq:conservation}
\end{equation}
This is verified as, 
we have for all $k \geq 1$ ($\lambda_k\neq 0$) by summing along
$\phi(s)$ in Eq.~\ref{eq:genvec1} and reminiscing Eq.~\ref{eq:tr_0},
\[
\begin{array}{l}
\displaystyle\sum_s\phi(s)(\bm{\mu}_{k,n}^+(s)  - \bm{\mu}_{k,n}^-(s)) \\\qquad\qquad=  -\frac{1}{\lambda_k}  \sum_s\phi(s)(\bm{\mu}_{k,n-1}^+(s)  - \bm{\mu}_{k,n-1}^-(s))\\
\displaystyle\qquad\qquad= \frac{(-1)^{n-1}}{\lambda_k^{n-1}}  \sum_s\phi(s)(\bm{\mu}_{k,1}^+(s)  - \bm{\mu}_{k,1}^-(s))\\
\displaystyle\qquad\qquad=  \frac{(-1)^{n-1}}{\lambda_k^{n}}\sum_{s'}\underbrace{\left[\sum_s\phi(s)(B^++B^-)_{ss'}\right]}_{=0 \ \text{see Eq.~\ref{eq:tr_0}}}(\bm{\mu}_{k,1}^+(s')  + \bm{\mu}_{k,1}^-(s')) = 0.
\end{array}
\]
And for $k=0$,
\[
\begin{aligned}
\sum_s\phi(s)\bm{\mu}_{0,2n}(s)  &= \sum_{s'}\underbrace{\left[\sum_s\phi(s)(B^++B^-)_{ss'}\right]}_{=0 \ \text{see Eq.~\ref{eq:tr_0}}}\bm{\mu}_{0,2n+1}(s').
\end{aligned}
\]
\subsection{Periodic boundary constraint}
\label{app:per-cond}
We now constrain the constants $(a_{\pm k,n})_{k,n}$ by the boundary
conditions. First, rewriting the general solution (\ref{eq:sol})
into,
\begin{equation}
\begin{array}{l}
\left(\begin{array}{c}
\bm{\gamma}^+\\
\bm{\gamma}^-\\
\end{array}\right)(r) \\=
\displaystyle\sum_{n=1}^{p_0}\sum_{l=n}^{p_0}\Big[\sum_{m=l}^{p_0}a_{0,m} \frac{\big(\tfrac{L}{2}\big)^{m-l}}{(m-l)!}\Big] \frac{\big(r-\tfrac{L}{2}\big)^{l-n}}{(l-n)!}
\left(\begin{array}{c}
\bm{\mu}_{0,n}   \\
(-1)^{n+1}\bm{\mu}_{0,n} \\
\end{array}\right)\\
    \displaystyle+ \sum_{k=1}^{d} \Big[ \exp\Big(-\lambda_k\Big(\frac{L}{2}-r\Big)\Big)\sum_{n=1}^{p_k}\sum_{l=n}^{p_k}\Big[\sum_{m=l}^{p_k}a_{k,m} \frac{\big(\tfrac{L}{2}\big)^{m-l}}{(m-l)!}\Big] \frac{\big(r-\tfrac{L}{2}\big)^{l-n}}{(l-n)!}
\left(\begin{array}{c}
\bm{\mu}_{k,n}^+   \\
\bm{\mu}_{k,n}^- \\
\end{array}\right)\\ \displaystyle+
 \exp\Big(\lambda_k\Big(\frac{L}{2}-r\Big)\Big)
 \sum_{n=1}^{p_k}\sum_{l=n}^{p_k}\Big[\sum_{m=l}^{p_k}(-1)^{n+1}a_{-k,m} \frac{\big(\tfrac{L}{2}\big)^{m-l}}{(m-l)!}\Big] \frac{\big(r-\tfrac{L}{2}\big)^{l-n}}{(l-n)!}
\left(\begin{array}{c}
\bm{\mu}_{k,n}^-   \\
\bm{\mu}_{k,n}^+ \\
\end{array}\right)\Big],     
\end{array}
\label{eq:sol_sym}  
\end{equation}
the periodic boundary condition
$ \bm{\gamma}^+(L/2^-)= \bm{\gamma}^-(L/2^-)$ yields,
\begin{equation}
\begin{array}{l}
 \displaystyle\sum_{n=1}^{p_0/2}2\Big[\sum_{m=2n}^{p_0}a_{0,m} \frac{\big(\tfrac{L}{2}\big)^{m-2n}}{(m-2n)!}\Big] \bm{\mu}_{0,2n} \\ \displaystyle\qquad\qquad+ \sum_{k=1}^{d}\sum_{n=1}^{p_k} \Big[\sum_{m=n}^{p_k}(a_{k,m}  -(-1)^{n+1}a_{k,m})\frac{\big(\tfrac{L}{2}\big)^{m-n}}{(m-n)!}\Big](\bm{\mu}_{k,n}^+ -    \bm{\mu}_{k,n}^-)=0.    
\end{array}
\label{eq:periodic}  
\end{equation}
As the vectors
$(\bm{\mu}_{k,n}^+-\bm{\mu}_{k,n}^-)_{k,n}, (\bm{\mu}_{0,2n})_n$ are
linearly independent, we obtain the condition,
\begin{equation*}  \def\arraystretch{1.4}
\left\{\begin{array}{l}
\sum_{m=2n}^{p_0}a_{0,m} \frac{(L/2)^{m-2n}}{(m-2n)!} = 0, \text{for all} \ n \leq  p_0/2\\
\sum_{m=n}^{p_k}(a_{k,m} - a_{-k,m} (-1)^{n+1})\frac{(L/2)^{m-n}}{(m-n)!} = 0, \text{for all} \ n \leq  p_k\\
\end{array}\right.,
\end{equation*}
constraining $\bm{\gamma}$ into the following symmetrical
decomposition in respect to
$\bm{\gamma}^+(r)$ and $\bm{\gamma}^-(L-r)$ (formally,
  when defined on the whole torus $[0,L]$),
\begin{equation}
\begin{array}{l}
\left(\begin{array}{c}
\bm{\gamma}^+(r)\\
\bm{\gamma}^-(r)\\
\end{array}\right) \\
\displaystyle\,\, = \sum_{l=0}^{p_0/2-1}b_{0,2l+1} \frac{\big(\tfrac{L}{2}-r\big)^{2l}}{(2l)!}
\left(\begin{array}{c}
\bm{\mu}_{\mathcal{B}}   \\
\bm{\mu}_{\mathcal{B}} \\
\end{array}\right) \\
 \displaystyle\,\,\,\,+
\sum_{n=1}^{p_0/2-1}\sum_{l=n}^{p_0/2-1}b_{0,2l+1} \frac{\big(\tfrac{L}{2}-r\big)^{2(l-n)}}{(2(l-n))!}\Big[
\left(\begin{array}{c}
\bm{\mu}_{0,2n+1}   \\
\bm{\mu}_{0,2n+1} \\
\end{array}\right)
-\frac{\big(\tfrac{L}{2}-r\big)}{2(l-n)+1}
\left(\begin{array}{c}
\bm{\mu}_{0,2n}   \\
-\bm{\mu}_{0,2n} \\
\end{array}\right)\Big]\\  
 \displaystyle \,\,\,\,   + \sum_{k=1}^{d}\sum_{n=1}^{p_k} \sum_{l=n}^{p_k}b_{k,l}\frac{\big(\tfrac{L}{2}-r\big)^{l-n}}{(l-n)!} \Big[ (-1)^{l-n} e^{-\lambda_k\left(\frac{L}{2}-r\right)}
\left(\begin{array}{c}
\bm{\mu}_{k,n}^
+   \\
\bm{\mu}_{k,n}^- \\
\end{array}\right)
+ e^{\lambda_k\left(\frac{L}{2}-r\right)}
\left(\begin{array}{c}
\bm{\mu}_{k,n}^-   \\
\bm{\mu}_{k,n}^+ \\
\end{array}\right)\Big] ,     
\end{array}
\label{eq:sol_expl}  
\end{equation} 
with,
\begin{equation*}
\def\arraystretch{1.4}
\left\{
\begin{array}{l}
b_{0,2l+1}=\sum_{m=2l+1}^{p_0-1}a_{0,m} \frac{\big(L/2\big)^{m-2l-1}}{(m-2l-1)!}, \ \text{for} \ 1 \leq 2l+1 \leq p_0-1,\\
a_{0,1}=0 \ \text{and} \ a_{0,2l}=-\sum_{m=2l+1}^{p_0}a_{0,m} \frac{\big(L/2\big)^{m-2l}}{(m-2l)!}, \ \text{for} \ 2 \leq 2l \leq p_0-2,\\
b_{k,l}=\sum_{m=l}^{p_k}a_{k,m} \frac{\big(L/2\big)^{m-l}}{(m-l)!}=(-1)^{l+1}\sum_{m=l}^{p_k}a_{-k,m} \frac{\big(L/2\big)^{m-l}}{(m-l)!}, \ \text{for} \ 1 \leq l \leq p_k.
\end{array}
\right. 
\end{equation*}
\subsection{Jamming boundary constraint}
\label{app:jam_cond}
We now consider the constraint imposed by the jamming interaction in
the explicit case of $K=Q$. The condition in $0^+$ set by $(C_{\mathcal{J}})$
leads to,
\begin{equation}
\left\{
\begin{array}{l}
w_{\mathcal{J}}B^+\bm{\mu}_{\mathcal{J}}^{-}
=   -\bm{\pi}_{\text{glob}}(0^+,-1)\\      
w_{\mathcal{J}}B^-\bm{\mu}_{\mathcal{J}}^{-}
=   \bm{\pi}_{\text{glob}}(0^+,1)
\end{array}
\right .,
\label{eq:0_plus_glob}
\end{equation}
with 
$$\bm{\mu}_{\mathcal{J}}^{-}=({\mu}_{\mathcal{J}}(-1,s))_s,$$
$$\bm{\pi}_{\text{glob}}(0^+,-1)=(\pi_{\text{glob}}(0^+,-1,s))_s,$$
 $$\bm{\pi}_{\text{glob}}(0^+,1)=(\pi_{\text{glob}}(0^+,1,s))_s.$$
It decomposes into,
\begin{equation}    
\left\{
\begin{array}{l}
\tfrac{w_{\mathcal{J}}}{w_{\mathcal{B}}}(B^++B^-)\bm{\mu}_{\mathcal{J}}^{-}
=  w_{\text{rel}}(\bm{\gamma}^+(0^+) -\bm{\gamma}^-(0^+))\\      
-\tfrac{w_{\mathcal{J}}}{w_{\mathcal{B}}}(B^+-B^-)\bm{\mu}_{\mathcal{J}}^{-}
=   (2w_{\text{eq}}\bm{\mu}_{\mathcal{B}} + w_{\text{rel}}(\bm{\gamma}^+(0^+) +\bm{\gamma}^-(0^+)))
\end{array}
\right ..
\label{eq:0_plus}
\end{equation}
It clearly appears that recovering a detailed-jamming solution
(i.e. $w_{\text{rel}}=0$) imposes
$(B^++B^-)\bm{\mu}_{\mathcal{J}}^{-}=0$,
i.e. $\bm{\mu}_{\mathcal{J}}^{-}\propto\bm{\mu}_{\mathcal{B}}$,
which, combined with the second condition, then makes the
detailed-jamming condition equivalent to $\bm{\mu}_{\mathcal{B}}$
being the unique eigenvector shared by $B^+$ and $B^-$ with
opposite eigenvalues. It also imposes that
$(B^+-B^-)\bm{\mu}_{\mathcal{J}}^{-}\propto\bm{\mu}_{\mathcal{B}}$,
i.e.  $\bm{\mu}_{\mathcal{J}}^{-}\propto\bm{\mu}_{0,2}$, which
leads to the condition
$\bm{\mu}_{0,2}\propto \bm{\mu}_{\mathcal{B}}$.

Now assuming $w_{\text{rel}}\neq 0$, i.e.
$\bm{\mu}_{\mathcal{J}}^{-}\not\propto\bm{\mu}_{\mathcal{B}}$ nor  $\bm{\mu}_{\mathcal{J}}^{-}\not\propto\bm{\mu}_{0,2}$, we
decompose $\bm{\mu}_{\mathcal{J}}^{-}$ along the two basis of
general eigenvectors,  
\begin{equation}
\begin{array}{rl}    
\bm{\mu}^-_{\mathcal{J}} &=
\sum_{n=1}^{p_0/2}c_{0,2n} 

\bm{\mu}_{0,2n}   
+ \sum_{k=1}^{d}  \sum_{n=1}^{p_k}c^-_{k,n}
\left(  \bm{\mu}_{k,n}^+ -  \bm{\mu}_{k,n}^-\right)\\
&=
\sum_{n=0}^{p_0/2-1}c_{0,2n+1} 

\bm{\mu}_{0,2n+1}   
+ \sum_{k=1}^{d}  \sum_{n=1}^{p_k}c^+_{k,n}
\left(  \bm{\mu}_{k,n}^+ +  \bm{\mu}_{k,n}^-\right)      
\end{array},
\label{eq:mu_j_decomp}
\end{equation}
so that,

\begin{equation*}
\begin{array}{rl}    \displaystyle
-(B^+-B^-)\bm{\mu}^-_{\mathcal{J}} & \displaystyle=
-\sum_{n=1}^{p_0/2}c_{0,2n} 

\bm{\mu}_{0,2n-1}   
\\& \displaystyle\quad- \sum_{k=1}^{d}  \Big[\lambda_kc^-_{k,p_k}
\left(  \bm{\mu}_{k,p_k}^+ +  \bm{\mu}_{k,p_k}^-\right)\\
& \displaystyle\quad\qquad\qquad+\sum_{n=1}^{p_k-1}(\lambda_kc^-_{k,n}+c^-_{k,n+1})
\left(  \bm{\mu}_{k,n}^+ +  \bm{\mu}_{k,n}^-\right)\Big],\\
 \displaystyle(B^++B^-)\bm{\mu}^-_{\mathcal{J}}  & \displaystyle=
\sum_{n=1}^{p_0/2-1}c_{0,2n+1} 

\bm{\mu}_{0,2n}   \\
& \displaystyle\quad+ \sum_{k=1}^{d}  \Big[\lambda_kc^+_{k,p_k}
\left(  \bm{\mu}_{k,p_k}^+ -  \bm{\mu}_{k,p_k}^-\right)
\\
&\quad\qquad\qquad +\sum_{n=1}^{p_k-1}(\lambda_kc^+_{k,n}+c^+_{k,n+1})
\left(  \bm{\mu}_{k,n}^+-  \bm{\mu}_{k,n}^-\right)\Big] .     
\end{array}
\end{equation*}
From the expression of $\bm{\gamma}^{\pm}$ (\ref{eq:sol_expl}), we get that,
\begin{equation*}
\begin{array}{rl}
&  \displaystyle
\bm{\gamma}^+(0)+
\bm{\gamma}^-(0) =
\sum_{n=0}^{p_0/2-1}\sum_{l=n}^{p_0/2-1}2b_{0,2l+1} \frac{\big(\tfrac{L}{2}\big)^{2(l-n)}}{(2(l-n))!}

\bm{\mu}_{0,2n+1}  
\\  
&  \displaystyle   + \sum_{k=1}^{d}\sum_{n=1}^{p_k} \sum_{l=n}^{p_k}b_{k,l}\frac{\big(\tfrac{L}{2}\big)^{l-n}}{(l-n)!} \Big[ e^{\left(\lambda_k\frac{L}{2}\right)}+(-1)^{l-n} e^{\left(-\lambda_k\frac{L}{2}\right)}\Big] 
\left(
\bm{\mu}_{k,n}^+  +
\bm{\mu}_{k,n}^- \right),     
\end{array}
\end{equation*}
and,
\begin{equation*}
\begin{array}{rl}
&  \displaystyle
\bm{\gamma}^+(0)-
\bm{\gamma}^-(0)

=  -\sum_{n=1}^{p_0/2-1}\sum_{l=n}^{p_0/2-1}2b_{0,2l+1} \frac{\big(\tfrac{L}{2}\big)^{2(l-n)+1}}{(2(l-n)+1)!}
\bm{\mu}_{0,2n}   \\ 
&  \displaystyle   + \sum_{k=1}^{d}\sum_{n=1}^{p_k} \sum_{l=n}^{p_k}b_{k,l}\frac{\big(\tfrac{L}{2}\big)^{l-n}}{(l-n)!} \Big[ -e^{\left(\lambda_k\frac{L}{2}\right)}+(-1)^{l-n} e^{\left(-\lambda_k\frac{L}{2}\right)}\Big]
\left(
\bm{\mu}_{k,n}^+   - \bm{\mu}_{k,n}^- \right).
\end{array}
\end{equation*} 
Plugging those expression in (\ref{eq:0_plus}), we obtain,
\begin{equation*}
\left\{ \begin{array}{l}
w_{\mathcal{J}}c_{0,2}=-2w_{\mathcal{B}}\big(w_{\text{eq}} +w_{\text{rel}}\sum_{l=0}^{p_0/2-1}b_{0,2l+1}\frac{(L/2)^{2l}}{(2l)!}\big)\\
w_{\mathcal{J}}c_{0,2(n+1)}=  -2w_{\mathcal{B}}w_{\text{rel}}\sum_{l=n}^{p_0/2-1}b_{0,2l+1}\frac{(L/2)^{2(l-n)}}{(2(l-n))!} , \ n\geq 1\\
w_{\mathcal{J}}c_{0,2n+1}=  -2w_{\mathcal{B}}w_{\text{rel}}\sum_{l=n}^{p_0/2-1}b_{0,2l+1}\frac{(L/2)^{2(l-n)+1}}{(2(l-n)+1)!} , \ n\geq 1\\
w_{\mathcal{J}}(\lambda_kc^-_{k,n} + c^-_{k,n+1})= -w_{\mathcal{B}}w_{\text{rel}} \sum_{l=n}^{p_k}b_{k,l}\frac{(L/2)^{l-n}}{(l-n)!} \Big[ e^{\lambda_k\frac{L}{2}}+(-1)^{l-n} e^{-\lambda_k\frac{L}{2}}\Big], \ n<p_k\\ 
w_{\mathcal{J}}(\lambda_k c^+_{k,n} + c^+_{k,n+1})= -w_{\mathcal{B}}w_{\text{rel}} \sum_{l=n}^{p_k}b_{k,l}\frac{(L/2)^{l-n}}{(l-n)!} \Big[ e^{\lambda_k\frac{L}{2}}-(-1)^{l-n} e^{-\lambda_k\frac{L}{2}}\Big], \ n<p_k\\
w_{\mathcal{J}}\lambda_kc^-_{k,p_k}= -w_{\mathcal{B}}w_{\text{rel}} b_{k,p_k} \Big[ e^{\lambda_k\frac{L}{2}}+ e^{-\lambda_k\frac{L}{2}}\Big]=  w_{\mathcal{J}}\lambda_k c^+_{k,p_k}\text{cotanh}\left(\lambda_k\frac{L}{2}\right)\\  
\end{array}\right..
\end{equation*}
Recovering a catenary-like solution (i.e. no polynomial modulation) then requires
\begin{equation}
\left\{ \begin{array}{l}
w_{\mathcal{J}}c_{0,2}=-2w_{\mathcal{B}}w_{\text{eq}}\\
c_{0,2(n+1)}=  c_{0,2n+1}=0 \ \text{for} \ n\geq 1 \\               
w_{\mathcal{J}}\lambda_kc^-_{k,1} = -w_{\mathcal{B}}w_{\text{rel}} b_{k,1} \Big[ \exp\left(\lambda_k\frac{L}{2}\right)+ \exp\left(-\lambda_k\frac{L}{2}\right)\Big]=  w_{\mathcal{J}}\lambda_k c^+_{k,1}\text{cotanh}\left(\lambda_k\frac{L}{2}\right)\\
c^-_{k,n} = c^+_{k,n}=b_{k,n}=0 \ \text{for} \ n > 1
\end{array}\right.,
\label{eq:catenary_jam}
\end{equation}
which, following (\ref{eq:mu_j_decomp}), is possible only if
the generalized eigenvector $\mu_{0,2}$ can be decomposed
over
$\big(\mu_{\mathcal{B}},(\mu_{k,1}^++\mu^-_{k,1})_{k\in
	N}\big)$, $N$ some subset of $\llbracket 1, d \rrbracket$,
as well as the corresponding vectors $(\mu_{k,1}^+-\mu^-_{k,1})_{k\in N}$
so that the third relation in (\ref{eq:catenary_jam}) is
obeyed.

More generally, the more vectors of
$\big(\mu_{\mathcal{B}},(\mu_{k,n}^++\mu^-_{k,n})_{k,n}\big)$
are needed to decompose $\mu_{0,2}$ over, the richer and less
close to its equilibrium counterpart the bulk relaxation
behavior will be. A special case is the detailed-jamming
situation where $\mu_{0,2}$ is proportional to
$\mu_{\mathcal{B}}$ and $N=\emptyset$ or the catenary-like
relaxation where only the eigenvectors are involved in the decomposition.

That analysis of the impact of the jamming boundary
constraint on the bulk relaxation can be generalized to a
tumble Markov kernel $K$ different from $Q$, leading to
different matrices $B_K^+,B_K^-$ in (\ref{eq:0_plus}) and the
consideration of their eigenvectors and generalized ones.

\end{document}